\documentclass[aps,pra,twocolumn,superscriptaddress]{revtex4-2}

\usepackage[english]{babel}
\usepackage{amsmath}
\usepackage{amsthm}
\usepackage{amssymb}
\usepackage{mathrsfs}
\usepackage{booktabs}
\usepackage{color}
\usepackage{hyperref}
\usepackage[ruled,vlined]{algorithm2e}
\usepackage{cases}
\usepackage{dblfloatfix}
\usepackage{multirow}
\usepackage{bbm}
\usepackage{enumitem}
\hypersetup{
  colorlinks   = true,  %Colours links instead of ugly boxes
  urlcolor     = blue,  %Colour for external hyperlinks
  linkcolor    = blue,  %Colour of internal links
  citecolor    = red    %Colour of citations
}
\usepackage{mathtools}
\usepackage{natbib}
\usepackage{pgfplots}
\usepackage{subfigure}
\usepackage{siunitx}
\usepackage{url}
\newcommand{\ket}[1]{ | #1 \rangle }

\newcommand{\rvline}{\hspace*{-\arraycolsep}\vline\hspace*{-\arraycolsep}}

\newcommand{\di}{{\rm d}}

\newcommand{\F}{{\sf F}}

\newcommand{\Y}{{\sf Y}}

\renewcommand{\S}{{\sf S}}

\mathtoolsset{showonlyrefs}

\theoremstyle{definition}

\theoremstyle{remark}

\theoremstyle{plain}
\newtheorem*{theorem*}{Theorem}

\usepackage{scalerel,stackengine}
\stackMath
\newcommand\reallywidehat[1]{%
\savestack{\tmpbox}{\stretchto{%
  \scaleto{%
    \scalerel*[\wi\di thof{\ensuremath{#1}}]{\kern-.6pt\bigwedge\kern-.6pt}%
    {\rule[-\textheight/2]{1ex}{\textheight}}%WIDTH-LIMITED BIG WEDGE
  }{\textheight}% 
}{0.5ex}}%
\stackon[1pt]{#1}{\tmpbox}%
}
\usepackage{comment}
\usepackage{float}

\SetArgSty{textnormal}

\makeatletter

\newenvironment{widetext2}{%
  \par\ignorespaces
  \setbox\widetext@top\vbox{%
   \vskip15\p@
   \hb@xt@\hsize{%
    \leaders\hrule\hfil
    \vrule\@height6\p@
   }%
   \vskip6\p@
  }%
  \setbox\widetext@bot\hb@xt@\hsize{%
    \vrule\@depth6\p@
    \leaders\hrule\hfil
  }%
  \onecolumngrid
  \let\set@footnotewidth\set@footnotewidth@ii
}{%
  \par
  \twocolumngrid\global\@ignoretrue
  \@endpetrue
}%

\makeatother

\begin{document}

\title{Fidelity Relations in an Array of Neutral Atom Qubits - Experimental Validation of Control Noise}

\author{D.A. \surname{Janse van Rensburg}} 
\altaffiliation[Corresponding author: ]{d.a.janse.van.rensburg@tue.nl}
\author{R.J.P.T. \surname{de Keijzer}}
\author{R.C. \surname{Venderbosch}}
\author{Y. \surname{van der Werf}}
\author{J.J. \surname{del Pozo Mellado}}
\author{R.S. \surname{Lous}}
\author{E.J.D. \surname{Vredenbregt}}
\author{S.J.J.M.F. \surname{Kokkelmans}}

\affiliation{Department of Applied Physics and Science Education, Eindhoven University of Technology, P. O. Box 513, 5600 MB Eindhoven, The Netherlands}
\affiliation{Eindhoven Hendrik Casimir Institute, Eindhoven University of Technology, P. O. Box 513, 5600 MB Eindhoven, The Netherlands}

\date{\today}

\begin{abstract}
Noise is a hindering factor for current-era quantum computers. In this study, we experimentally validate the theoretical relationships between amplitude noise of the control signal and qubit state fidelity. The experiment comprises a 10$\times$10 site optical tweezer array stochastically loaded with single rubidium-85 atoms. A global microwave field is used to manipulate the state of the hyperfine qubits. With precise control of the time-dependent amplitude of the microwave drive, we apply control signals featuring artificial noise. We systematically analyze the impact of various noise profiles on the fidelity distribution of the quantum states. The measured fidelities are compared against theoretical predictions made using the stochastic Schr{\"o}dinger equation. Our results show a good agreement between the experimentally measured and theoretically predicted results. This validation is consequential, as the model provides critical information on noise identification and optimal control protocols in NISQ-era quantum systems.  
\end{abstract}

\maketitle

\section{Introduction}

Quantum computing promises revolutionary results in various fields by solving complex problems that are intractable for modern-day classical computers \cite{qcMonroe, nisq}. Among the different qubit architectures suggested, neutral-atom quantum systems have garnered significant attention due to their scalability and long coherence times. Alkali atoms have been widely used in atomic physics research, and single-atom control in optical tweezers has enabled advances in quantum computing and simulation, as demonstrated in e.g. Refs. \cite{bluvstein_logical_2024,PhysRevLett.132.263601,graham_multi-qubit_2022}. Current-day systems are stepping stones to universal quantum computers, particularly because of limitations regarding noise influences and achievable qubit numbers. An important challenge for quantum hardware is maintaining high fidelity of quantum states in the presence of control noise. 

Recent interest has been sparked in the development of control pulses robust to several types of noise \cite{robust1,robust2,robust3,madhav}. In these works, noise is treated as a static offset from the ideal value (i.e. low-frequency noise), and thus is not time dependent on the duration of the pulse. An alternative approach is to leverage the Lindblad equation to design pulses that optimize the fidelity of the mean output state \cite{lindblad1,lindblad2}. In Ref.~\cite{fvqoc}, time-dependent noise is treated and pulses that are robust for every individual preparation of an output state (instead of only the mean) are developed using the stochastic Schr\"{o}dinger equation (SSE) \cite{colorednoisepaper}. Here, Ref.~\cite{fvqoc} uses the analytical control noise vs.\ qubit fidelity predictions from Ref.~\cite{ssetheory}.

In this paper, we show the match between the theoretical models of Ref.~\cite{ssetheory} and experimental realities. This is achieved by experimentally verifying the predicted impact that different types of colored noise on the amplitude of the control signal have on the fidelity of quantum states. In our experiment, we validate the fidelity noise models on a neutral atom qubit microwave transition. As the fidelity noise models of Ref.~\cite{ssetheory} are qubit architecture agnostic, our results provide substantial evidence for their validity on other architectures such as optical transitions in ion traps, microwaves in transmon qubits, voltages in charge or topological qubits, and others.

\begin{figure}
    \centering
    \includegraphics[width=0.86\linewidth]{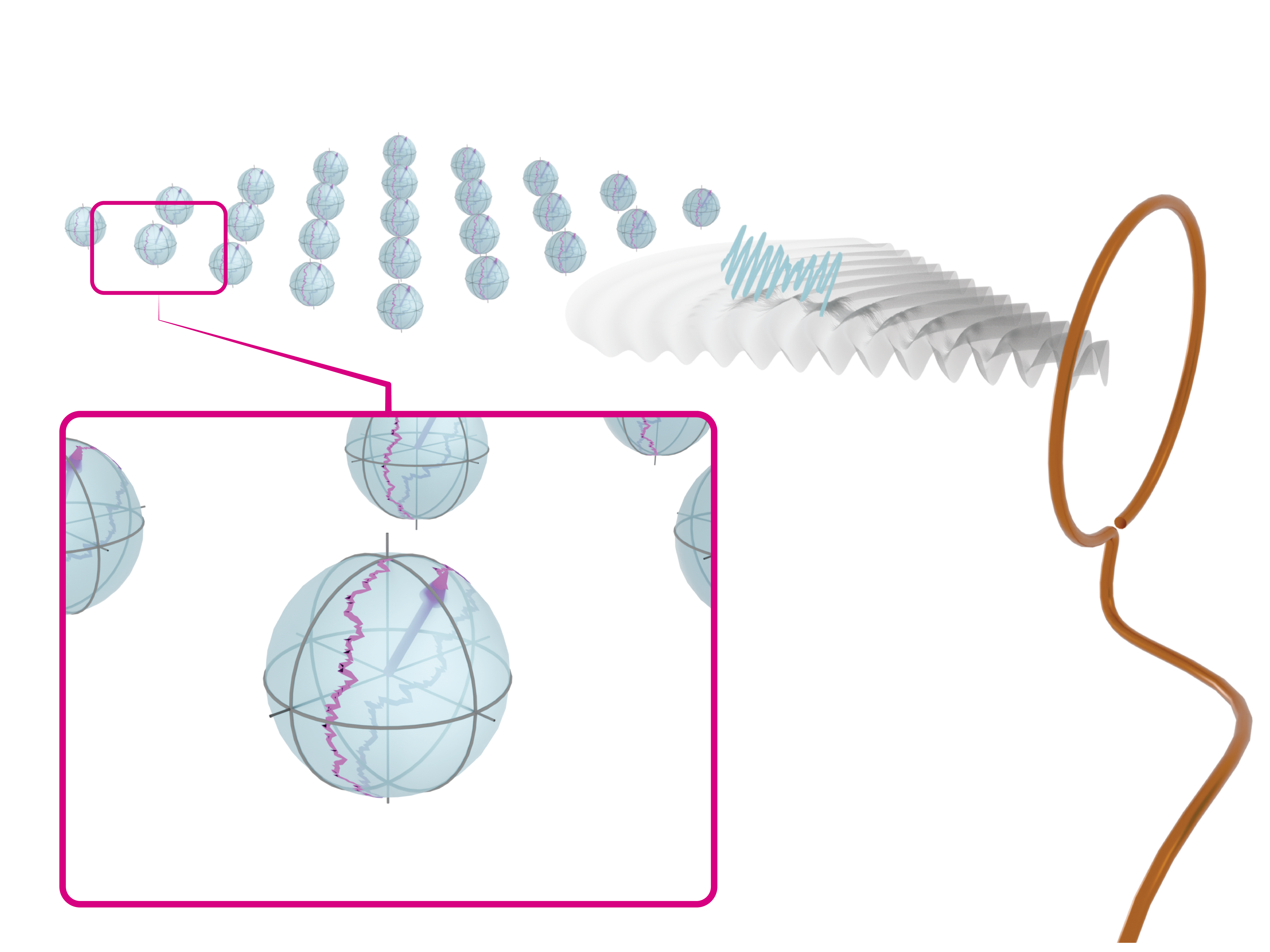}
    \caption{Conceptual representation of the experiment. Using a microwave antenna, a control signal with superimposed noise is sent to an array of neutral atom qubits. The state evolution under the applied noise leads to a reduction of the final state fidelity, as depicted by the pink arrow.}
    \label{fig: graphical abstract}
\end{figure}

Our setup encodes the qubit state on the hyperfine clock states of single Rb-85 atoms, which are trapped and cooled in an array of optical tweezers. The qubits can be prepared in an initial state and are coherently manipulated by a microwave field. Using precise pulse generation, artificial noisy control signals can be systematically introduced to the qubits. The corresponding output states can subsequently be measured, and due to the 100 optical tweezers that are all addressed by a global microwave pulse, statistics on these states are rapidly collected, as illustrated in Fig.~\ref{fig: graphical abstract}. Moreover, as all trapped atoms experience the same noise, each experiment is effectively repeated approximately 300$\times$50 times (accounting for a $\approx0.5$ loading probability) for the same noise realization These two aspects make the microwave transition on Rb-85 hyperfine qubits an ideal platform to validate the noise fidelity models. The results are then compared to outcomes obtained under ideal, non-noisy conditions. This comparison allows for a quantification of the degradation in the accuracy of the final quantum state due to control noise, i.e., the fidelity. By repeatedly injecting the same realization of artificial noise, statistics can be gathered on the fidelity of that specific realization. Doing this for multiple realizations allows for the calculation of the entire distribution of fidelities. 

Several other works have investigated the influence of noise on quantum computer performance. Our method is unique in determining the full distribution of states instead of only the density matrix, and its ability to handle control noise varying on the timescale of a single pulse. Ref.~\cite{noisecheck1} simulates the influence of dephasing, while Ref.~\cite{noisecheck2} benchmarks quantum computing performance against a noise model where Pauli gates are randomly inserted in the circuit and the state is described on a density matrix level. Other methods use randomized circuit benchmarking \cite{randomizedtheory} to characterize their fidelity performance \cite{randomized1,randomized2,randomized3}.

In Sec.~\ref{Section:SSEEqn}, we briefly discuss the relevant analytic results of Ref.~\cite{ssetheory}, which are experimentally verified in this work. In subsequent sections (Secs.~\ref{section:Experiment},\ref{sec:ExperimentalVerification}), we provide a detailed description of our experimental setup, including the methodologies for generating and introducing noise on control pulses and the techniques used for fidelity measurement. Our results (Sec.~\ref{sec:results}) show good agreement with the theoretical models, demonstrating clear correlations between the characteristics of control noise and the fidelity of quantum states. These findings enhance our understanding of noise influence and can serve as guidelines or benchmarks for more reliable quantum computing technologies.

\section{Stochastic Schr\"{o}dinger equation}
\label{Section:SSEEqn}
 Single-qubit control for neutral atoms is described by the control Hamiltonian $H(t)$ given by
\begin{equation}
\label{eq:hamiltonian}
    H(t)=\frac{1}{2}\sum_j \Omega_j(t) \sigma_x^j +\Delta_j(t)(I-\sigma_z^j),
\end{equation}

where $\sigma_x^j$ and $\sigma^j_z$ are Pauli matrices, $\Omega_j$ and $\Delta_j$ are the Rabi frequency and detuning at site $j$, respectively controlled by the intensity and the frequency of the driving electromagnetic field. Noiseless quantum states $\phi$ \footnote{In this work, we choose to refrain from using standard bra-ket notation for readability purposes.} evolve under this Hamiltonian as
\begin{equation}\label{eq:se}
    \di \phi=-iH(t)\phi\, \di t, \quad \phi(0)=\phi_0.
\end{equation}
However, noise is ubiquitous in the current noisy intermediate-scale quantum (NISQ) era \cite{nisq} and should be taken into account when looking at the evolution of states. 

The evolution of a stochastic quantum state $\psi$ evolving under classical control noise $X$ is described by the Stochastic Schr\"{o}dinger equation (SSE)
\begin{equation}
\label{eq:sse}
	\di \psi=-iH\psi\, \di t-\frac{1}{2} S^\dagger S \psi\, \di [X]_t - i S\psi\, \di X_t,
\end{equation}
where $S$ is the noise operator. Here, $X$ can be any semi-martingale noise \cite{semimartingale}, including white noise (WN), Ornstein-Uhlenbeck (OU) noise, or Brownian motion noise (BM), and $\di[X]_t$ is the quadratic variation of this noise \cite{ito2014stochastic}. The SSE serves as a physically interpretable unraveling of the Lindblad master equation \cite{unraveling}, where the classical nature of the noise is justified by the semi-classical atom-light interactions valid at high photon number \cite{classical1,classical2}. 

This work analyzes the control noise on the intensity, resulting in the noise operator $S=\sigma_x^j$.  We are interested in the dynamic behavior of the accuracy of our system, which we quantify using the probability distribution of fidelities $F$ defined as $F=|\phi^\dagger\psi|^2.$ In Ref.~\cite{ssetheory}, predictions of qubit fidelities for various control noise realizations are derived for a large class of possible noise operators. For example, for noise operators $S$ obeying $S^\dagger S=I$ (such as $S=\sigma_x^j$),
\begin{equation}
\label{eq:fidelity}
\begin{aligned}
    \F &=\frac{1}{2}(1+\S_0^2)+\frac{1}{2}(1-\S_0^2)\cos(2(X_t-X_0)) \\
    &=\cos^2(X_t-X_0) + \S_0^2\sin^2(X_t-X_0),
\end{aligned}
\end{equation}
with $\S_0=\phi_0^\dagger S\phi_0\in\mathbb{R}$. From this distribution, expressions for the mean fidelity $\mathbb{E}[F]$ and higher moments can be derived, as given in App.~\ref{app:analexpres}. Note that this relation holds for any semi-martingale $X_t$ and moments of this noise can vary highly with the type of noise. 

Verifying these theoretical noise-fidelity correspondence models on actual experimental systems is necessary to ensure their validity and applicability across different research setups. Consequently, these validated models offer a tested benchmark that other experiments can use to evaluate whether their specific setups meet certain required criteria by inferring expected fidelities from their noise spectra. In this work, we verify the above distribution of fidelities for various noise profiles.

\section{Rubidium-85 tweezer platform}
\label{section:Experiment}
Our setup is based on trapped single Rb-85 atoms in an array of red-detuned optical tweezers using light derived from an \SI{813}{\nano\meter} laser. 
Single atoms are loaded from a three-dimensional magneto-optical trap (3D MOT) inside a glass vacuum cell into an array of 10$\times$10 optical tweezers generated using a spatial light modulator (SLM) and focused through a microscope objective. Following the loading process, achieving single-site loading probabilities of $\approx$ 0.5, polarization-gradient cooling is applied and adiabatic reduction of the trapping potential to further cool the atoms and increase the coherence time. Afterward, a uniform bias magnetic field of \SI{5.66}{G} is applied to define the quantization axis. 

Qubits are encoded in the magnetically insensitive hyperfine ground states $\ket{0} =$~5$^2$S$_{1/2}$~$\ket{F=2, m_F=0}$ and $\ket{1}=$~5$^2$S$_{1/2}$~$\ket{F=3,m_F=0}$. Coherent control is achieved using global microwaves pulses, denoted $R_{\phi}^G(\theta)$, resonant at \SI{3.035}{\giga\hertz}. State preparation in $\ket{0}$ is performed by alternating optical and microwave pulses, as described in App.~\ref{app:PrepDetect}. We infer a temperature of \SI{12}{\micro\kelvin} after the state preparation procedure using the release and recapture technique \cite{RRTemp}. State selective measurements are performed by inducing losses of $\ket{1}$ atoms from the trap using a light pulse; the remaining atoms are detected by a subsequent fluorescence image on the $5S_{1/2} \ket{F=3}$ $\leftrightarrow$ $5P_{3/2} \ket{F'=4}$ transition.

In our setup, we measure the qubit relaxation time $T_1$ = \SI{205\pm3}{\milli\second} and the inhomogeneously broadened dephasing time $T_2^*$ = \SI{2.01\pm0.02}{\milli\second} for the ensemble  average of the array.
We attribute the primary sources of decoherence to differential light shifts and off-resonant scattering from the \SI{813}{\nano\meter} trapping light acting on the hyperfine qubits. For individual array sites, we measure $\langle T_2^*\rangle=$\SI{3.5\pm0.5}{\milli\second} indicating that inhomogeneous differential light shifts caused by trap depth inhomogeniety lead to a faster dephasing of the ensemble average. Details of the setup and the characterization measurements are further described in App.~\ref{app:PrepDetect}. 

\section{Experimental Verification}
\label{sec:ExperimentalVerification}
\subsection{Experimental Implementation of Artificial Noise}
\label{sec:NoiseImp}
\begin{figure}
    \centering
    \includegraphics[width=0.9\linewidth]{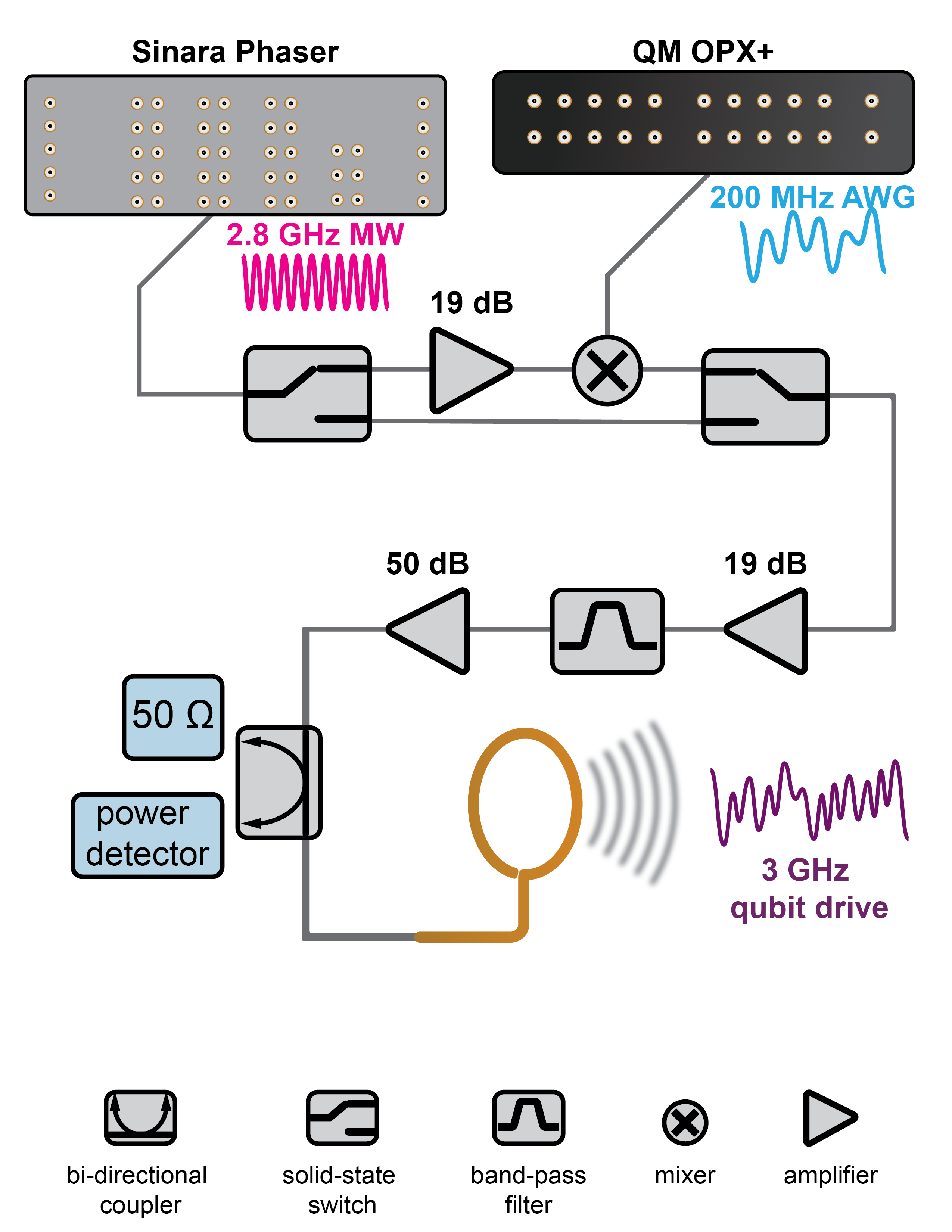}
    \caption{Simplified diagram of the microwave drive circuit. The artificial noise is introduced as a time varying amplitude of the \SI{200}{\mega\hertz} produced by a Quantum Machines Operator-X module (QM OPX+). This signal is upconverted to \SI{3.035}{\giga\hertz} to be resonant with the atomic transition, by mixing with a ~2.8 GHz base frequency from a Sinara Phaser, controlled by Artiq.}
    \label{fig:MWDiagram}
\end{figure}

For the global microwaves which drive the qubit transition, an agile, low-noise, and high-stability microwave source with fast modulation of amplitude is utilized in our experiments. This microwave signal is derived from mixing a \SI{2.8035}{\giga\hertz} oscillator output originating from a Sinara Phaser, which is part of the Artiq control system \cite{bourdeauducq_2016_51303} for the experiment, with a Quantum Machines OPX (AWG) at \SI{0.2}{\giga\hertz}. The \SI{2.8035}{\giga\hertz} carrier and lower frequency mixing product are filtered out using a cavity band-pass filter (Mini-Circuits ZVBP-3100A-S+). Both the Phaser and QM outputs are referenced to a \SI{10}{\mega\hertz} oscillator, disseminated by the Dutch National Metrological Institute (VSL) via the White Rabbit protocol \cite{WhiteRabbit}. The operation mode of the microwave source can be changed to utilize only the Sinara Phaser at \SI{3.035}{\giga\hertz} for simpler tasks, such as state preparation, via an RF switch. Using a pre-amplifier in series with a \SI{100}{\watt} amplifier (Mini-Circuits ZHL-100W-352+), the microwave signal is sent to a home-built loop antenna located a few millimeters from the glass cell, with which a maximum Rabi frequency of $\approx2\pi \times\SI{160}{\kilo\hertz}$ can be achieved. A fraction of the power sent to the antenna is tapped off using a directional coupler and is monitored for diagnostic purposes using a Mini-Circuits PWR-8PW-RC power sensor. To provide a reference for the microwave system performance at zero added noise, we performed single-qubit randomized benchmarking at a Rabi frequency $\Omega=2\pi\cdot \SI{117}{\kilo\hertz}$ and found a Clifford gate fidelity $F_C = 0.999653(5)$ (see App.~\ref{app: rb}).  A schematic representation of the microwave drive circuit is shown in Fig.~\ref{fig:MWDiagram}.

The microwave pulse sent to the atoms (without artificial noise) has a Rabi frequency of $2\pi\times\SI{50}{\kilo\hertz}$  and a duration of \SI{200}{\micro\second}. This pulse induces a 20$\pi$ rotation around the $x$-axis. We run an automated calibration routine before and after collecting every data point to ensure that slow drifts in the microwave amplitude during extended data collection periods are compensated for. This calibration routine sets a scaling factor to the base MW amplitude to ensure that the \SI{200}{\micro\second} pulse (without added noise) has a pulse area of 20$\pi$. 

The artificial noise is applied by modulating the amplitude of the \SI{200}{\mega\hertz} signal from the AWG which results in the modulation in the amplitude of the microwave pulse sent to the atom array. Although the model from Ref.~\cite{ssetheory} is generally applicable to amplitude, detuning, and phase noise, this work focuses on the implementation of amplitude noise. A further study implementing detuning noise would require signal generation capabilities and diagnostics beyond what is currently implemented. The amplitude is varied at a time step $\Delta t$ =\SI{1}{\micro\second}, the motivation for this time step is described in App.~\ref{app:ampresp} and App.~\ref{app:stepsize}. 
The artificial noise is applied for various durations $t$, which we define as the segment of a constant \SI{200}{\micro\second} pulse. A noiseless pulse, corresponding to $t=0$, would be a constant-valued \SI{200}{\micro\second} pulse, ensuring an even number of $\pi$-pulses. A $t=$\SI{100}{\micro\second} pulse, has artificial noises added for its first \SI{100}{\micro\second}, after which the last \SI{100}{\micro\second} are constant-valued. 
\subsection{Simulation model}

To simulate the dynamics of the fidelity distribution, we first generate $N_r$ noise profiles $X_i$ ($i=1,\ldots,N_r$) for white noise (WN), Ornstein-Uhlenbeck noise (OU), and Brownian motion (BM). The details of the noise generation are given in App.~\ref{app:stepsize}. Using a weak second-order Platen scheme (see App.~\ref{app:stochint}), we integrate the stochastic Schr\"{o}dinger equation Eq.~\eqref{eq:sse} for a \SI{200}{\micro\second} pulse encoding 20 consecutive $\pi$-pulses at a Rabi frequency of $\Omega/2\pi=\SI{50}{\kilo\hertz}$ as well as with addition of the generated noises. This results in a set of noisy states $\psi_{ij}$, with $j$ indicating the site index. In this integration, we take into account a correction for the non-linear Rabi frequency versus amplitude calibration (See App.~\ref{app:ampresp}), and inhomogeneity across the atom array (see App.~\ref{app:rabisinhomogeneity}). This yields for each noise profile $X_i$ and site $j$ a fidelity $F_{ij}=|\psi_{ij}^\dagger\phi|^2\in[0,1]$.

For each of these realizations, we need to consider the measurement statistics. An experimental measurement will only show a decisive result when an atom is present in the trap, which happens with probability $p_{c}\in[0,1]$, caused mostly by the $\approx0.5$ loading rate as well as atom loss. Furthermore, there are state preparation and measurement (SPAM) errors $p_{0\rightarrow 1}$ and $p_{1\rightarrow 0}$ for measuring a 1 when the outcome should have been 0 and vice versa, respectively. Both of these phenomena are assumed to be site-independent. In total, there are $N_{m}\cdot N_a$ measurement outcomes. The simulation model for SPAM errors is detailed in App.~\ref{app:spam}.

The probabilities $p_c, p_{0\rightarrow 1}$ and $ p_{1\rightarrow 0}$ are extracted from the experiment resulting in simulated measurement fidelity variables $F_{m,i}$, see App.~\ref{app:spam}. From the distribution of Eq.~\eqref{eq:simulatedmeasure}, the uncertainties of both the simulation and the experimental results can be extracted. 

\section{Results}
\label{sec:results}

To verify the theoretical noise-fidelity models, we measure the qubit fidelity for three types of amplitude noise, as described in App.~\ref{app:stepsize}. In each experiment, we generate the noises and implement them experimentally as well as in simulation via Eq.~\eqref{eq:sse}. We compare these results to the analytic expressions of Eqs.~\eqref{eq:expectations} and ~\eqref{eq:variances}. In all experiments, the values of $\gamma$ (the noise strength) and $\kappa$ (the noise damping rate) are chosen so that the injected noise profiles are the dominant error source and the resulting fidelity graphs clearly show the characteristics of the noise profiles.

\begin{figure}
    \centering
    \includegraphics[width=0.95\linewidth]{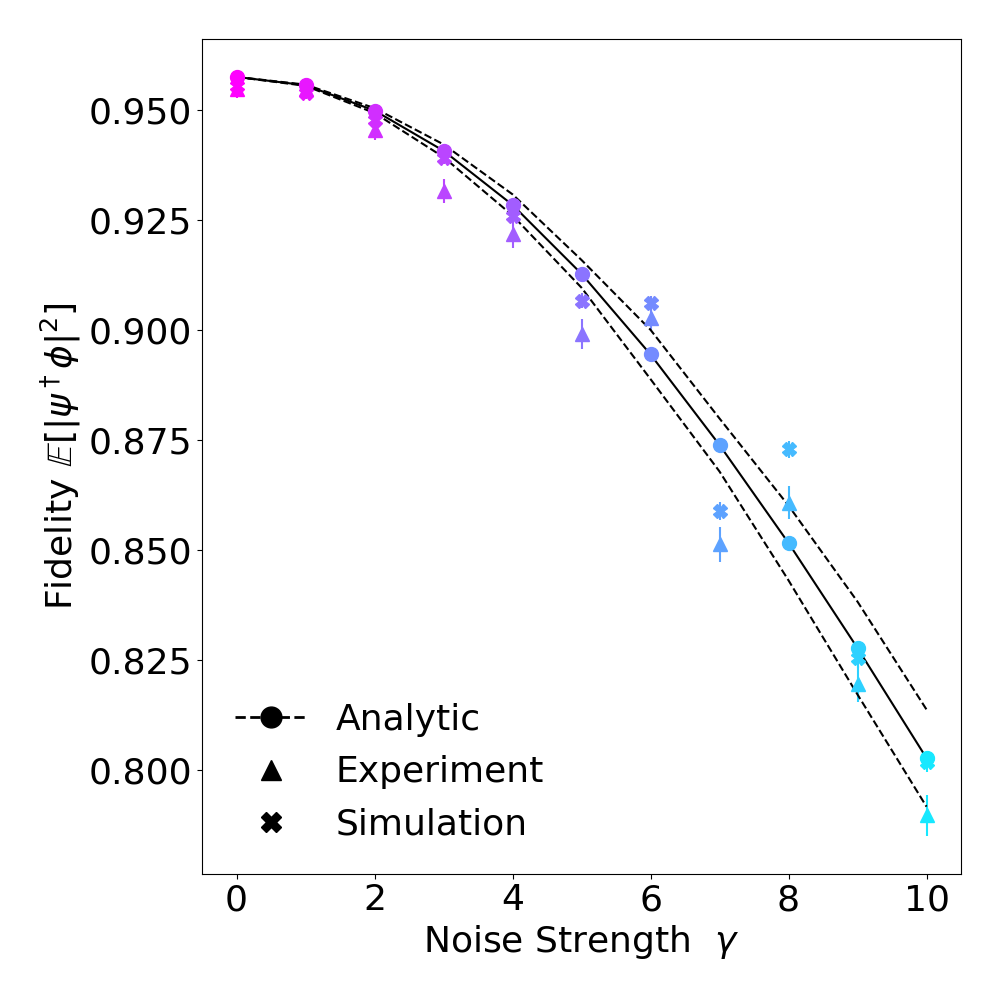}
    \caption{Qubit fidelity for various Ornstein-Uhlenbeck noise strengths. Average final fidelities after \SI{200}{\micro\second} pulse over 75 realizations, 100 sites, and 300 measurements per realization. Analytic results (circles) with 1$\sigma$ deviation over the possible realizations (dashed lines), experimental (triangles) and simulation (squares) for varying noise strengths $\gamma$ with error bars indicating 1$\sigma$ over the realizations. Ornstein-Uhlenbeck noise with 200 time steps and $\kappa=5\cdot 10^3$  s$^{-1}$.}
    \label{fig:gammas}
\end{figure}

The first experiment analyzes the fidelities of the qubit array for OU noise with various noise strengths $\gamma$. Figure~\ref{fig:gammas} shows the result, where the dotted lines surrounding the average are the values indicating $1\sigma$ deviations from the sample mean when considering 75 realizations (as estimated by simulation). From Fig.~\ref{fig:gammas}, a good correspondence between the analytic, experimental and simulation results can be observed. %Deviations can be caused by miscalibrations or experimental fluctuations, as discussed below.\\

\begin{figure}
    \centering
    \includegraphics[width=0.90\linewidth]{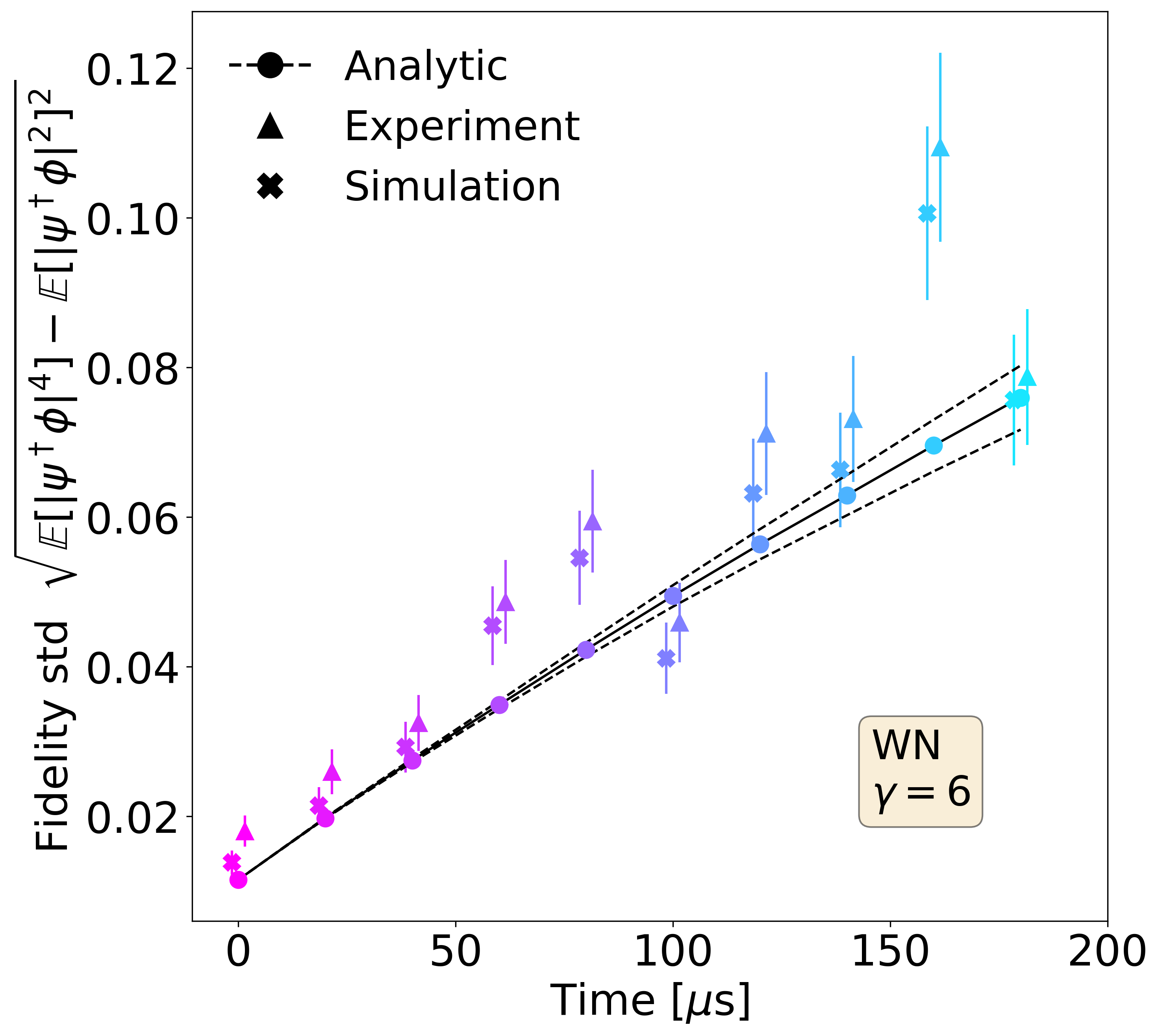}
    \caption{Qubit fidelity standard deviation for white noise as a function of time. Average final fidelities after \SI{200}{\micro\second} pulse over 75 realizations, 100 sites, and 300 measurements per realization. Analytic results (circles) with 1$\sigma$ deviation over the possible realizations (dashed lines), experimental (triangles) and simulation (squares) with error bars indicating 1$\sigma$ over the realizations. White noise with 200 time steps and $\gamma=6$  s$^{-1/2}$.}
    \label{fig:variances}
\end{figure}

Fig.~\ref{fig:fidelityaverages} shows the average fidelity for the three different noise profiles: white noise (WN), Ornstein-Uhlenbeck noise (OU), and Brownian motion (BM). The horizontal axis represents the noise duration $t$. In these results, the choice of the maximum $t$ is arbitrary and does not have any physical significance. There is a good agreement between the experimental results, numerical simulations, and analytical predictions, all of which follow similar trends that reflect the characteristic behavior of each noise type, i.e. the linear reduction in fidelity for WN, the damping effect inherent in OU and the accelerating decrease in fidelity characteristic of BM representing a random walk, see App.~\ref{app:stepsize} for more details and power spectral densities (PSDs) of these noises. Unlike Lindblad based models, the SSE model allows for the prediction of higher moments of the fidelities, for instance for white noise as shown in Fig.~\ref{fig:variances}. This knowledge can be useful for error-correction applications \cite{error_correction}, where one might want to ensure that all preparations reach a certain fidelity, rather than the average. Again good agreement is seen between the analytic predictions, simulations and experiment.

In Fig.~\ref{fig:fidelitydistributions}, the full underlying distributions (histograms) for the data points in Fig.~\ref{fig:fidelityaverages} are given and show good correspondence between the kernel density estimates \cite{chen2017tutorial} of the simulation results (thick lines) and the experimental data. For WN, OU and BM, the tails of the kernel density simulation capture the relevant behavior of the real experimental data. The simulations at $t=0$, corresponding to zero noise, are equivalent for all types of noise in Fig.~\ref{fig:fidelitydistributions}. This deviation between simulation and experiment is caused by the finite number of measurements. % For $t>0$, deviations are caused by different noise realizations sampled from the same distribution and. 
A mismatch between simulation and experiment is observed at high fidelities between 0.95 and 1 for white noise in Fig.~\ref{fig:fidelitydistributions}a at $t=\SI{180}{\micro\second}$. This is partially explained by the spread in amplitudes, which is higher for WN compared to OU or BM. BM features the smallest amplitude jumps of the noise traces and shows better overlap at high fidelity. For visibility reasons, only a single typical error bar is shown in the $x$-direction. As each fidelity measurement is the average result of $300$ repetitions on an average of $50$ atoms, the uncertainty of its mean is $\sqrt{F(1-F)/N}$ with $N=300\times50$. We choose to indicate the $x$-error around the peaks of the distribution resulting in an error of $\approx 0.002$, which is approximately the bin size. Note that the errors over the entire histogram mostly average out as high sample fidelities from low base fidelities cancel with low sample fidelities from high base fidelities. The error bar indicates the error for a single count, but these errors will generally average out over all counts.      

Despite the good agreement between experiment, simulation, and theory, we observe that the experimental fidelities are typically lower than those from theory and simulation. This may potentially be due to additional real-world noise sources and calibration errors not taken into account in the models. We provide further information on these potential errors and the mitigation strategies we have implemented to reduce their influence in App~\ref{app: noise sources}. 

\begin{widetext2}
    \begin{minipage}[b]{\linewidth}
        \begin{figure}[H]
            \centering
            \includegraphics[width=0.90\linewidth]{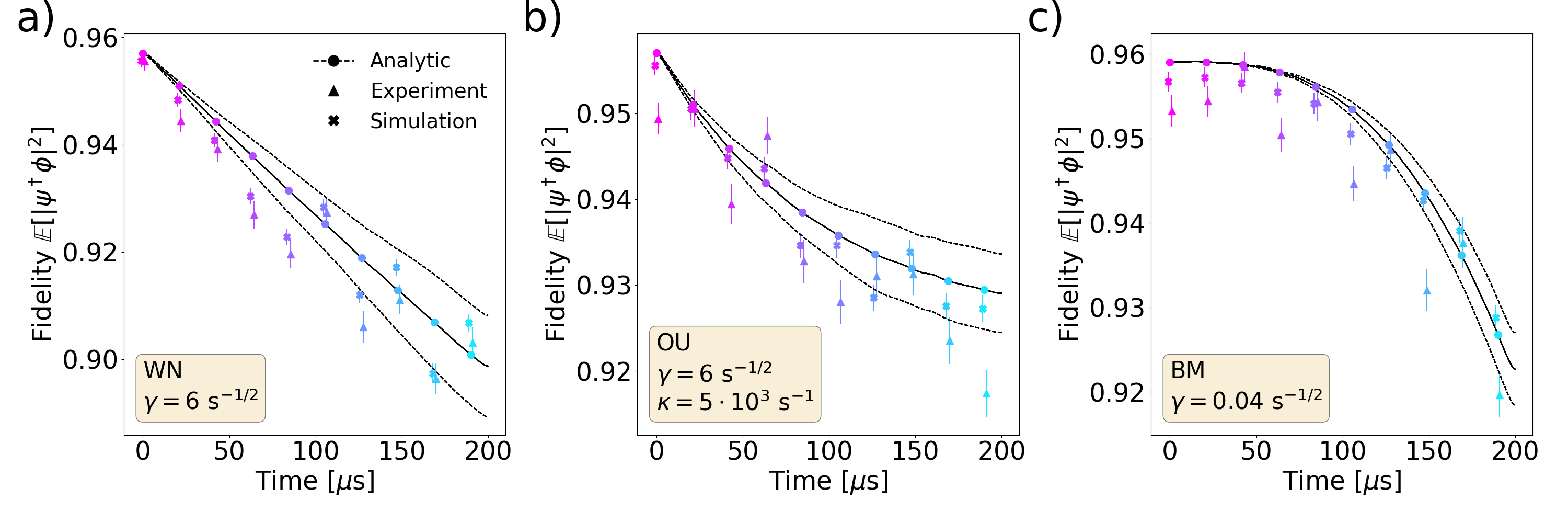}
    \caption{Noise evolution over time for 3 types of noise. Average fidelities over 75 noisy realizations, 100 sites, and 300 measurements per realization for equivalent \SI{200}{\micro\second} pulses in experiment (triangles) and simulation (squares). Analytic results (black) together with their $1\sigma$ standard deviation over 75 realizations (dotted lines). a) white noise (WN) with $\gamma=6$  s$^{-1/2}$, b) Ornstein-Uhlenbeck noise (OU) with $\gamma=6$ s$^{-1/2}$ and $\kappa=5\cdot 10^3$ s$^{-1}$, and c) Brownian motion (BM) with $\gamma=0.04$ s$^{-1/2}$. Error bars indicating 1$\sigma$ over the realizations}
            \label{fig:fidelityaverages}
        \end{figure}    
    \end{minipage}
\end{widetext2}

\begin{widetext2}
    \begin{minipage}[b]{\linewidth}
        \begin{figure}[H]
            \centering
            \includegraphics[scale=0.24]{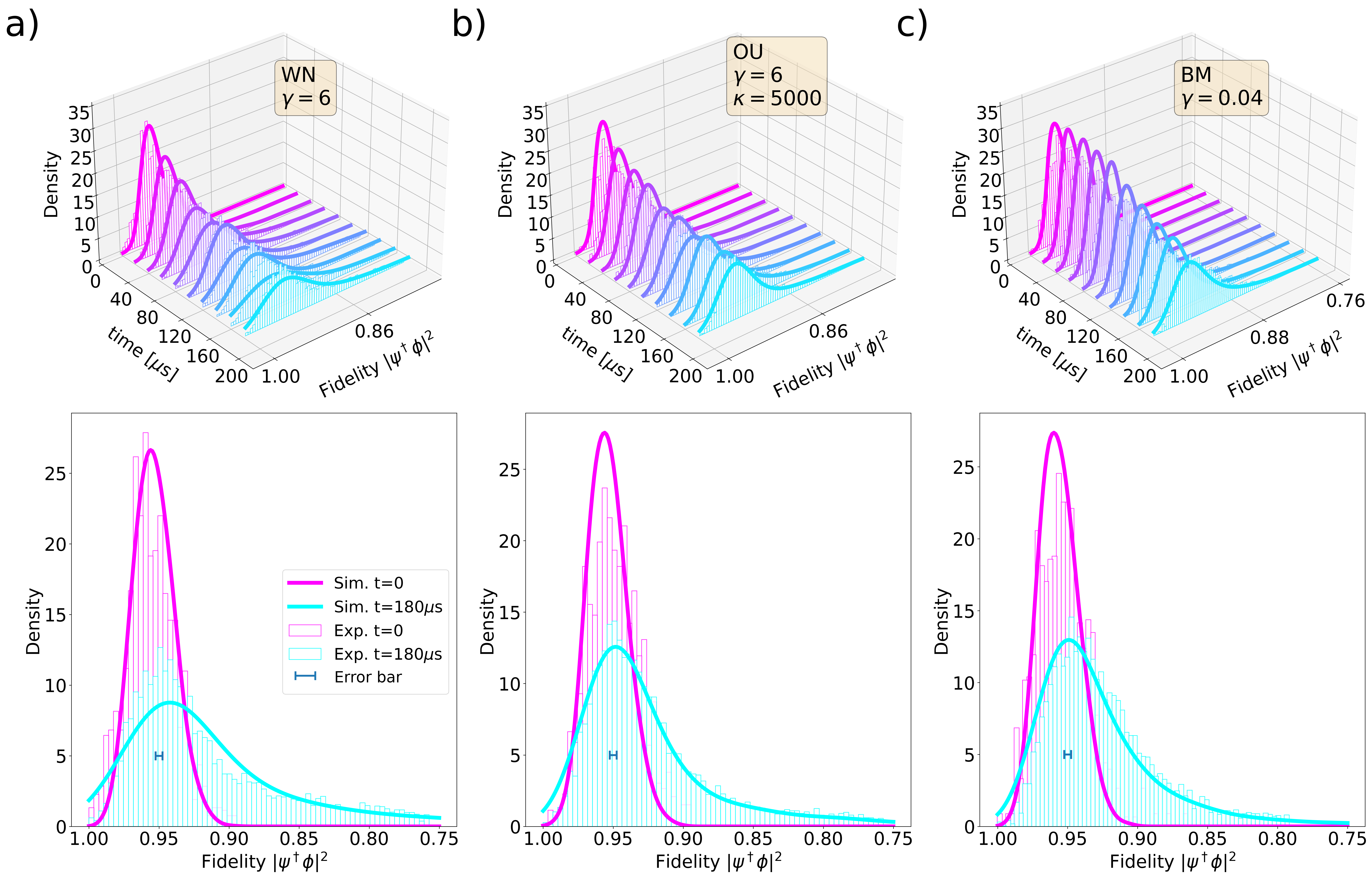}
    \caption{Measurements of fidelity distributions for various types of noise.Fidelity distributions over 75 noisy realizations, 100 sites, and 300 measurements per realization for equivalent \SI{200}{\micro\second} pulses in experiment (histograms) and kernel density estimates of simulation (solid lines). a) white noise (WN) with $\gamma=6$  s$^{-1/2}$, b) Ornstein-Uhlenbeck noise (OU) with $\gamma=6$  s$^{-1/2}$ and $\kappa=5\cdot10^3$  s$^{-1}$, and c) Brownian motion (BM) with $\gamma=0.04$  s$^{-1/2}$. Bottom figures show simulation and experimental distributions at both $t=0$ and $t=\SI{180}{\micro\second}$. Note the inverted $x$-axis. Error bar in $x$-direction indicates typical $x$-error $\approx 0.002$ for sample mean of fidelity measurement, which is distributed as a binomial variable $B(F,N)$ with $N\approx300\times 50$ being the average number of measurements per realization.}
            \label{fig:fidelitydistributions}
        \end{figure}    
    \end{minipage}
\end{widetext2}

\section*{DISCUSSION}
In this work, we experimentally verify results on the dependence of the fidelity on noise from Ref.~\cite{ssetheory}, obtained using stochastic Schrödinger equations (SSE) for various noise profiles. Noise profiles are generated and implemented on top of the microwave intensity profile that addresses the $|0\rangle\leftrightarrow|1\rangle$ transition in an array of Rb-85 hyperfine qubits. By repeating noise realizations and measuring on all 100 available array sites, statistics for a noise profile can be gathered rapidly, allowing for the experimental resolution of the fidelity distributions. These distributions and their characteristics are compared to simulation and analytic results based on the same noise realizations. 

In general, good agreement is seen between the experimental, simulation and analytic results, verifying the fidelity relations as proposed by Ref.~\cite{ssetheory}. The validation of the model showcases the use of the stochastic Schr\"{o}dinger equation as a valuable tool in identifying noise sources and predicting their influences on control line performance in quantum systems. This has potential utility in predicting fidelities for well-characterized noise sources or diagnosing noise sources from the measured fidelity distributions. The full fidelity distributions will give more information for this process.With the identification of the type of noise present on the control signal, quantum optimal control schemes that can consider the control noise when calculating the optimum trajectory along the qubit state space may result in higher fidelity state preparation \cite{F-VQOC}. \\

The noise strengths considered in this work are larger than would be typically encountered in an experiment. We chose to operate in this regime to amplify the effect of the artificial noise above the contributions inherent in the experiment. The analytic work in Ref.~\cite{ssetheory} does not make assumptions about the strength of the noise and we therefore expect these results will be applicable to more realistic noise strengths encountered in a typical experiment. Furthermore, as the predicted relations in Ref.~\cite{ssetheory} are qubit architecture-independent, our results provide evidence for the validity of these models for control lines over many platforms. 

As an extension of this work, which focuses purely on amplitude noise, one could verify more noise fidelity relations. For instance, noise on the detuning of the microwaves, or non-commuting noise resulting from mixing of both detuning and intensity contributions, would constitute an interesting study to simulate laser addressing of optical qubit transitions. Such experiments will require the capability to program pulses with high-bandwidth detuning control and diagnostics. Moreover, we would like to extend the noise models beyond the three models discussed in this work and Ref.~\cite{ssetheory} to include more experimentally realistic cases. Finally, the results would benefit from gathering data on the survival probabilities and SPAM errors per site instead of an overall number.

\section*{ACKNOWLEDGEMENTS}
We thank Jasper Postema, Raul F. Santos, Madhav Mohan, and Jasper van de Kraats for fruitful discussions. We thank the SrMic team at the University of Amsterdam and the Sr quantum computing team at Eindhoven University of Technology for discussions during the construction of the experimental setup. This research is financially supported by the Dutch Ministry of Economic Affairs and Climate Policy (EZK), as part of the Quantum Delta NL program,  the Horizon Europe programme HORIZON-CL4-2021-DIGITAL-EMERGING-01-30 via the project 101070144 (EuRyQa), and by the Netherlands Organisation for Scientific Research (NWO) under Grant No.\ 680.92.18.05. 

\section*{AUTHOR CONTRIBUTIONS}
D.J.v.R., Y.v.d.W, R.V. and R.d.K conceptualized the work, performed the data analysis, interpreted the results and drafted the manuscript; D.J.v.R., Y.v.d.W, R.V. and J.d.P.M designed, constructed and characterized the experimental setup and collected the data. R.d.K performed the simulations. E.V., R.L. and S.K. supervised the work. All authors reviewed and edited the manuscript. 
\section*{COMPETING INTERESTS}
The authors declare no competing interests.

\section*{DATA AND CODE AVAILABILITY}
The data and code that support the findings of this study have been uploaded to GitLab \footnote{All data and code used are available publicly from \url{https://gitlab.tue.nl/s1658271/sse.git}}. Additional data is available from the corresponding author upon reasonable request.

\newpage
\bibliography{Bibliography.bib}

\newpage
\textcolor{white}{.}
\newpage
\newpage
\section*{APPENDICES}
%\subsection{Optical tweezers}

\subsection{Qubit Preparation and Detection}
\label{app:PrepDetect}
\begin{figure}[b]
    \centering
    \includegraphics[width=1.0\linewidth]{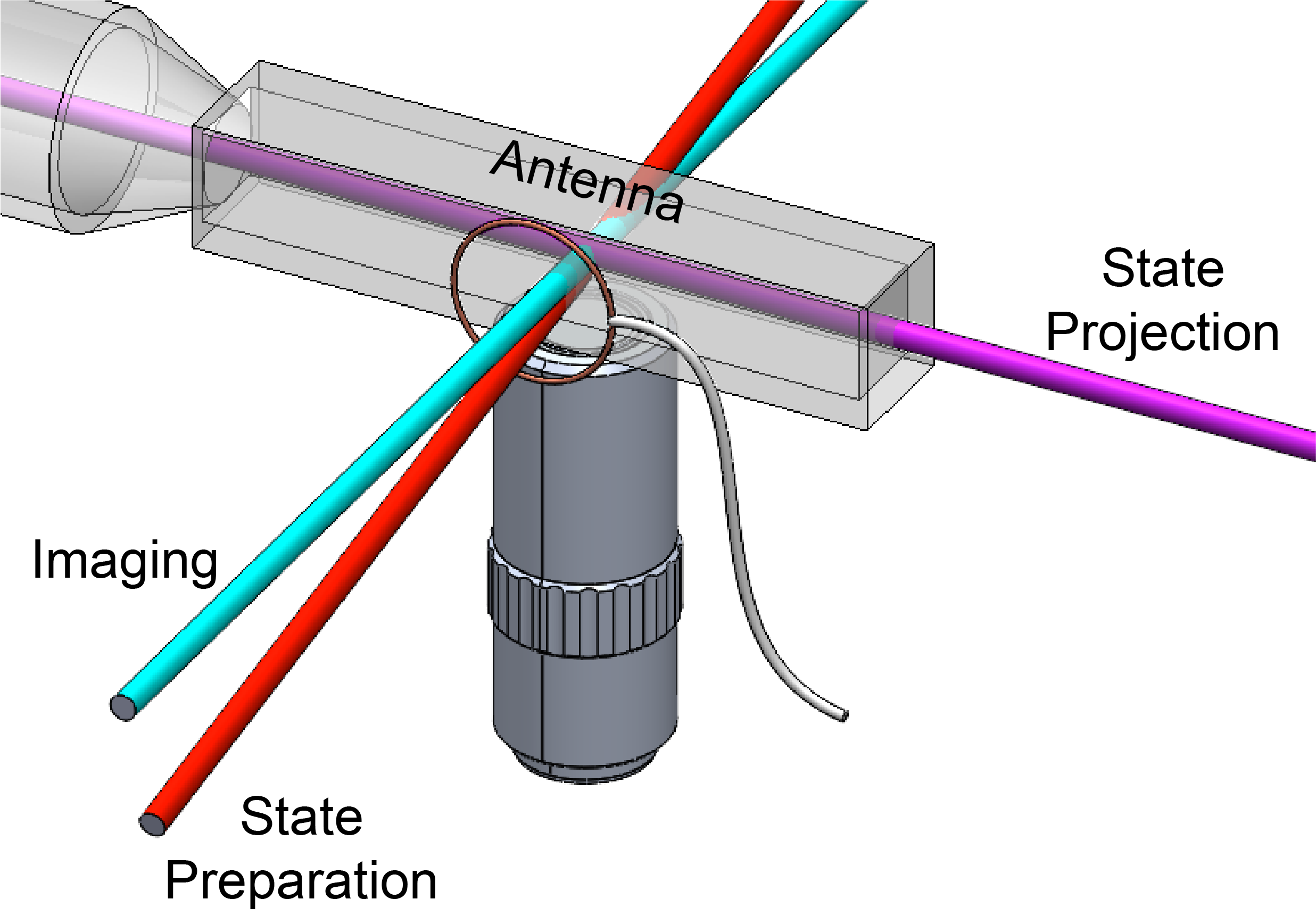}
    \caption{Simplified diagram of the experimental setup surrounding the UHV glass cell. The tweezer array is formed in the focus of the microscope objective where the 3 depicted laser beams, which are used for imaging, state preparation and state projection overlap. The 3 beams are all derived from a single \SI{780}{\nano\meter} laser and the different colors of the beams are merely for visualization purposes. The magnetic field coils surrounding the glass cell are not depicted. The microwave antenna is placed near the surface of the glass cell, approximately \SI{1.7}{\centi\meter} from the tweezer array. }
    \label{fig:StatePrepDiagram}
\end{figure}

\begin{figure}
    \centering
    \includegraphics[width=1.0\linewidth]{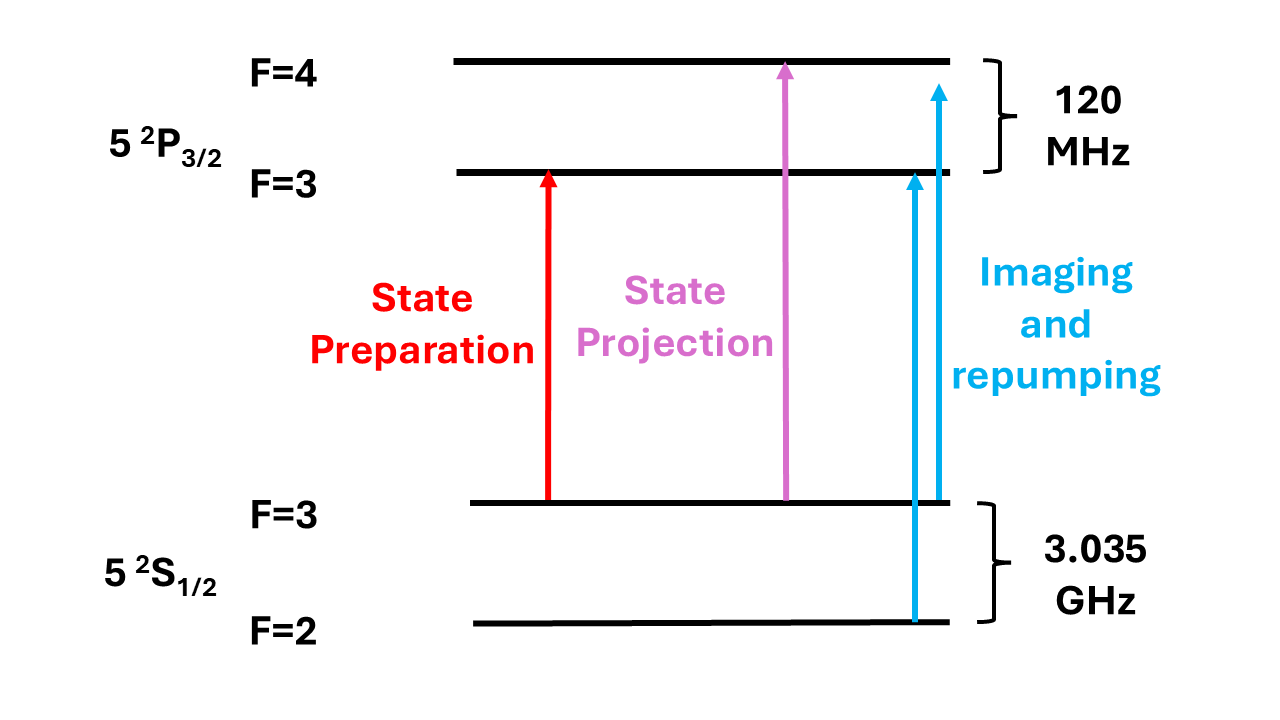}
    \caption{Simplified diagram of the relevant level structure of $^{85}$Rb. Transitions addressed for imaging, state preparation and state projection are indicated by the colored arrows. Energies are not to scale. }
    \label{fig:LevelStruct}
\end{figure}

Initially, atoms are loaded into the optical tweezers from a cold atomic cloud that originates from a magneto-optical trap. Once the atoms are loaded in the optical trap, counter-propagating beams in one spatial axis are used to induce fluorescence of the atoms in tweezers and perform polarization-gradient cooling on the 5$^2$S$_{1/2}$~$\ket{F = 3}$~$\leftrightarrow$~5$^{2}$P$_{3/2}$~$\ket{F' = 4}$ transition whilst repumping on the 5$^2$S$_{1/2}$~$\ket{F = 2}$~$\leftrightarrow$~5$^{2}$P$_{3/2}$~$\ket{F' = 3}$ transition. The single spatial axis is chosen to reduce scattered light on the EMCCD camera sensor. The geometric configuration of the laser beams for imaging, state preparation and state projection is shown in Fig.~\ref{fig:StatePrepDiagram} and the relevant level structure of Rb-85 is depicted in Fig.~\ref{fig:LevelStruct}. 

\begin{figure}[b]
    \centering
    \includegraphics[width=1.0\linewidth]{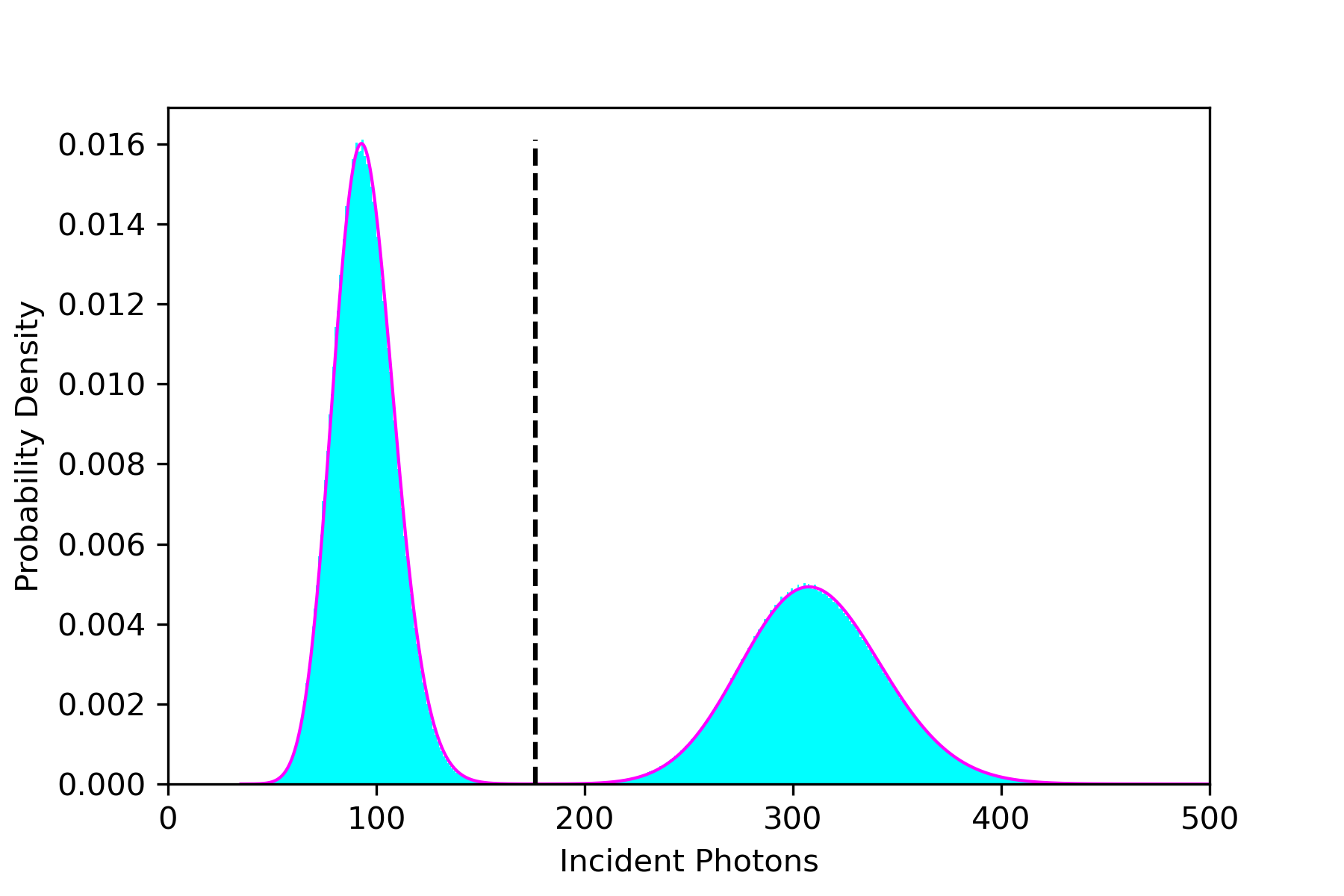}
    \caption{Imaging histogram for all ROIs combined for the 841500 loading cycles comprising the main dataset. The threshold for distinguishing between the zero-atom and single-atom distributions is indicated by the vertical dashed line. }
    \label{fig:Histogram}
\end{figure}

The presence of an atom in a tweezer is determined via a fluorescence image. If the amount of photons detected in a region of interest (ROI) exceeds a threshold value, an atom is determined to be in the tweezer. Figure~\ref{fig:Histogram} shows a histogram of fluorescent photon counts which shows the bimodal structure characteristic of optical tweezers loaded in the collisional blockade regime\cite{PhysRevLett.89.023005}. With a threshold value of 149 photons (indicated by the dashed line in Fig.~\ref{fig:Histogram}) we determine via the method in \cite{ImageFidelity} an array averaged imaging fidelity of $0.999$ with a 0.001 standard deviation for different array sites.

The loading and imaging are performed at a trap depth of $U_0/k_B \approx$ \SI{0.71}{\milli\kelvin}. The tweezers are adiabatically ramped down to a depth of $U_0/k_B \approx$ \SI{0.13}{\milli\kelvin} for state preparation, microwave manipulation, and state projection. After state projection, the trap depths are increased to their initial value for the final image.  
For qubit state preparation, we make use of alternating laser and microwave pulses to prepare the atoms in the $5S_{1/2} \ket{F=2,m_F=0} := \ket{0}$ state. A bias field of \SI{5.66}{G} is applied, and the atoms are coarsely pumped into the $5S_{1/2} \ket{F=2}$ state using resonant $\sigma^+ - \sigma^-$ polarized light on the $5S_{1/2} \ket{F=3}$ $\leftrightarrow$ $5P_{3/2} \ket{F'=3}$ transition. Then, four microwave $\pi$-pulses are applied in sequence to transfer any population from $5S_{1/2} \ket{F=2,m_F=i}$ to $5S_{1/2} \ket{F'=3,m_F'=i}$ for $i \in \{-2,-1,2,2\}$. This optical pumping process is repeated 35 times to reach a $\ket{0}$ state preparation fidelity which combined with the detection fidelity results in a $\approx$~0.94 $\pi$-pulse contrast. With the release and recapture method \cite{RRTemp}, a typical atom temperature of \SI{12}{\micro\kelvin} is measured after the state preparation procedure.

For state projection, a resonant $\sigma^+$ polarized laser beam is used to pump the $\ket{1}:=\ket{F=3,m_F=0}$ atoms to the $\ket{F=3,m_F=3}$ stretched state where the repeated scattering of photons on the closed 5 $^2$S$_{1/2}$ $\ket{F=3,m_F=3}$ $\leftrightarrow$ 5 $^2$P$_{3/2}$ $\ket{F=4,m_F=4}$ transition leads to loss of atoms from the tweezer. This process leaves the $\ket{0} := \ket{F=2,m_F=0}$ atoms unaffected. This allows for the mapping of the survival of an atom, as measured in a subsequent fluorescence image, to $\ket{0}$ and the loss of the atom to $\ket{1}$. 
\subsection{Randomized benchmarking}
\label{app: rb}
\begin{figure}
    \centering
    \includegraphics[width=0.95\linewidth]{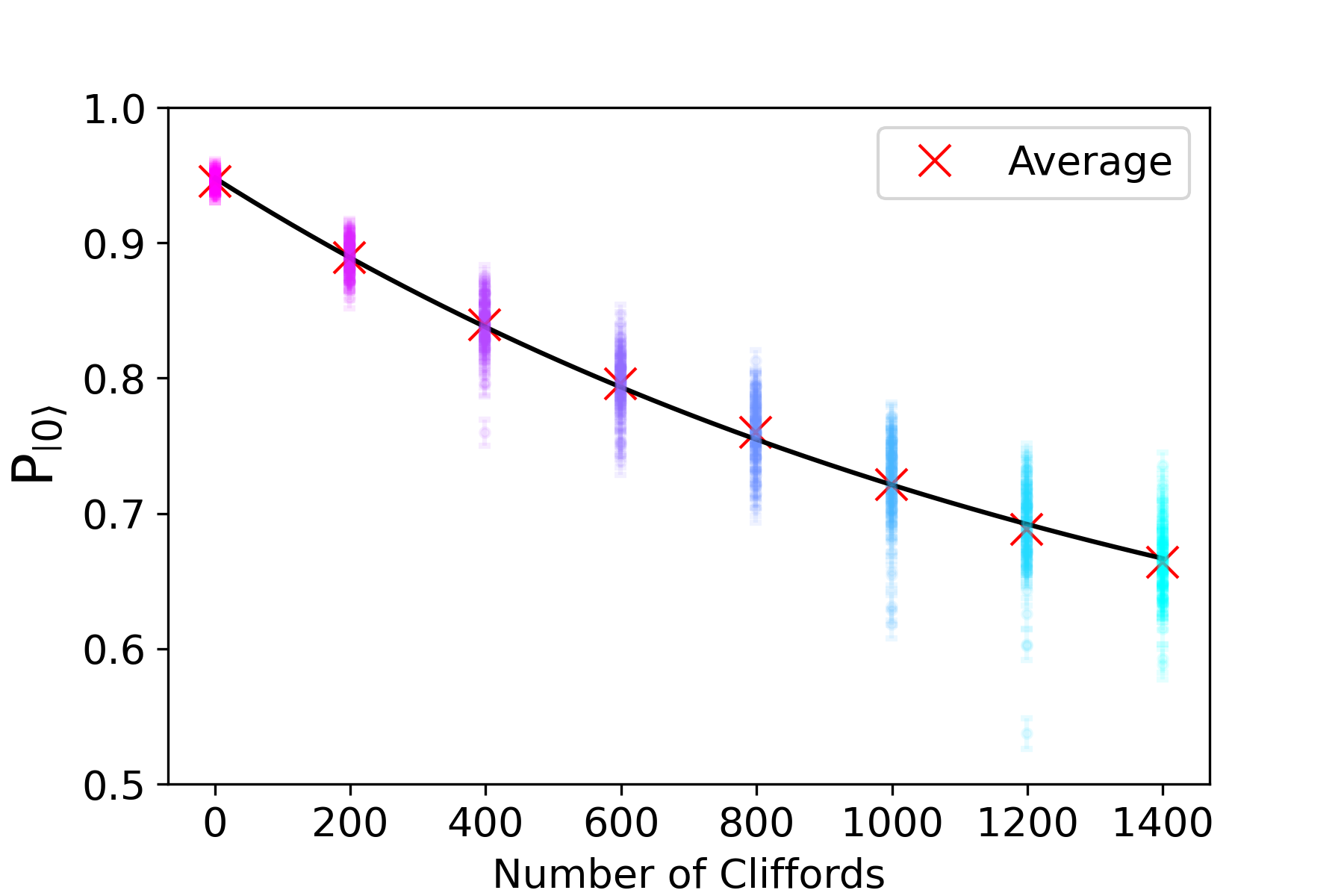}
    \caption{Global single-qubit randomized benchmarking using full sampling with 75 random sequences per Clifford length, represented by the coloured markers, that are each averaged over 75 measurements. The decay of the return probability to the $\ket{0}$ state is fitted to extract a Clifford fidelity $F_C = 0.999653(5)$.}
    \label{fig:rb}
\end{figure}

To investigate single-qubit gate performance in the limit of zero added noise, we perform randomized benchmarking using gates drawn from the Clifford group $C_1$, followed by an inversion gate \cite{randomized1, randomized2}. In our implementation, global $R_x$ and $R_y$ rotations are realized by means of microwave pulses with the rotation axis of each pulse selected by shifting the phase of the microwave source, while $R_z$ rotations are implemented by decomposing them into a sequence of $R_x$ and $R_y$ operations. Due to spatial variations in the Rabi frequency across the atomic array, we benefit from using composite pulses to suppress pulse-length errors. Specifically, we use SCROFULOUS pulses \cite{SCROFULOUS} to construct Clifford gates, yielding an average pulse area of $4.46\pi$ per gate. The results of the benchmarking experiment are shown in Fig.~\ref{fig:rb}. We fit the decay of the return probability to state $\ket{0}$ using the expression $\frac{1}{2}+\frac{1}{2}(1-d_0)(1-d)^n,$ where $n$ is the number of Cliffords, $d_0$ accounts for SPAM errors, and the Clifford gate fidelity is given by $F_C = 1-d/2$ \cite{randomized4}. The final Clifford gate fidelity is $F_C = 0.999653(5)$ whilst operating at a Rabi frequency $\Omega = 2\pi \cdot \SI{117}{\kilo\hertz}$ with $d_0=0.104(5)$. This gate fidelity serves as a performance metric for the microwave system without added noise, and as an independent method to assess SPAM errors, complementing the approach detailed in App.~\ref{app:spam}.

\subsection{Error sources}
\label{app: noise sources}

The experimental fidelities are typically lower than those from theory and simulation, which is likely due to additional real-world noise sources and potential calibration errors not accounted for in the models. There are several factors that can contribute to this reduced experimental fidelity: Rabi frequency inhomogeneity; survival probability and SPAM errors; amplitude instabilities in the microwave power; rise time for generating the amplitude noise traces; intensity noise in the optical tweezer laser; microwave phase noise; phase flicker from the spatial light modulator (SLM); and magnetic field stability. 

\subsubsection{Rabi Inhomogeneity}
\label{app:rabisinhomogeneity}
\begin{figure}
    \centering
    \includegraphics[width=1.0\linewidth]{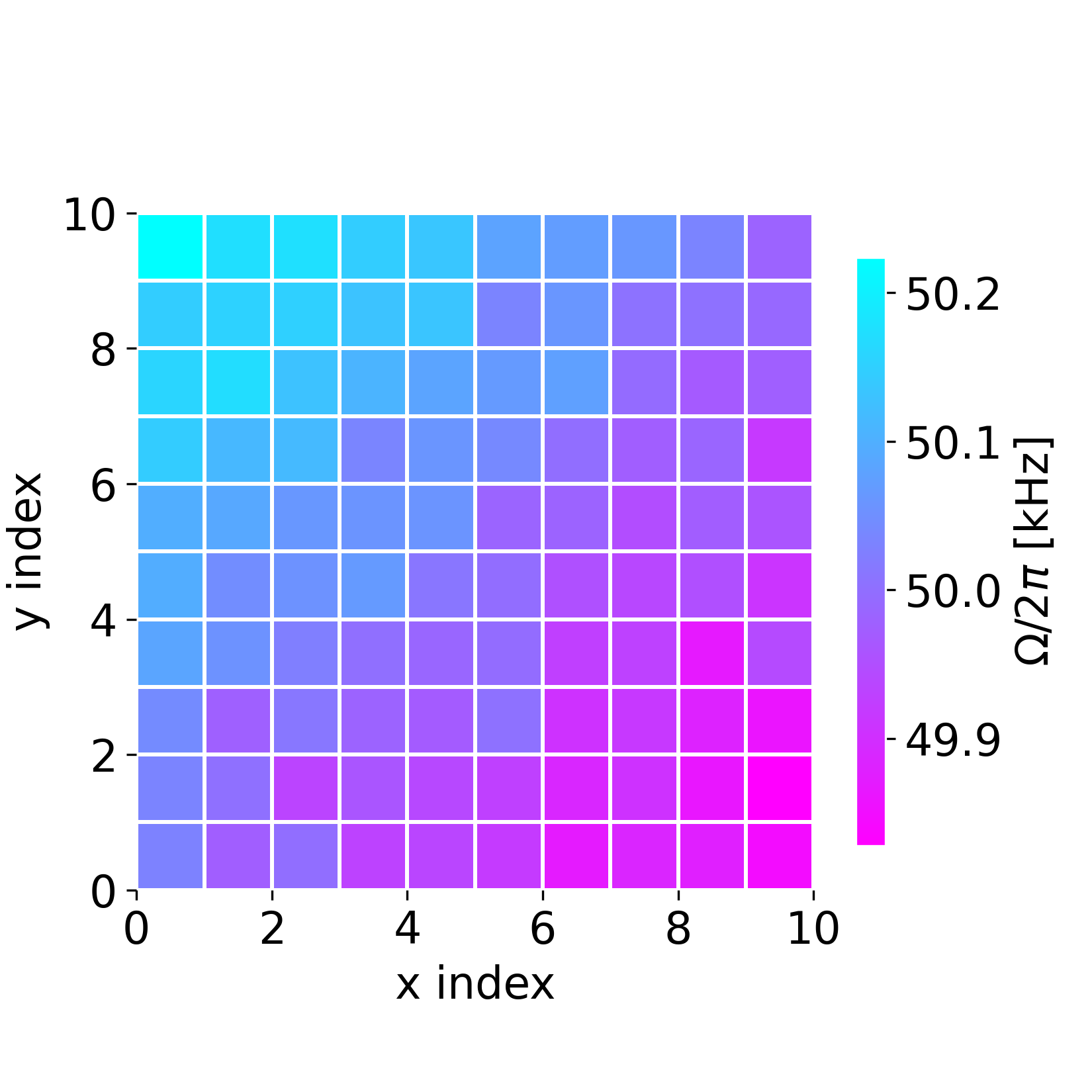}
    \caption{Measured Rabi frequencies at different array sites for a global microwave pulse. The deviation in the Rabi frequency for different sites is due to a spatially non-uniform microwave field. The inhomogeneity is taken into account during the data analysis.}
    \label{fig:RabiGradient}
\end{figure}

The amplitude of the microwave field is not uniform over the array. This leads to each array site experiencing a different Rabi frequency, as depicted in Fig.~\ref{fig:RabiGradient}. We characterize the site-dependent Rabi frequency by fitting the pulse-length-dependent oscillation in the probability of detecting the $\ket{0}$ state ($P_{\ket{0}}$) for each array site. We calibrate the microwave amplitude such that the array-averaged Rabi frequency equals $2\pi \times$\SI{50}{\kilo\hertz}. The inhomogeneity of the Rabi frequency over the array is characterized by a coefficient of variation of 0.14 \%. The Rabi frequency for each site is calculated using the calibrated scaling factor relative to the mean of the array. 
%Picture of Rabi oscillations at 150 kHz and at 50 kHz.
%\begin{figure}
%    \centering
%    \includegraphics[width=1.0\linewidth]{pictures/rabimicrowaves.png}
%    \caption{Caption}
%    \label{fig:enter-label}
%\end{figure}

% \subsubsection{Phase noise}
% We inadvertently collected the data with excess 20 kHz phase noise on the 3 GHz microwave signal from a badly operating phase locked loop (PLL) internal in the microwave signal generator. This caused a large resonant dephasing for Rabi frequencies around 20 kHz. This is visible in a spin lock measurement with X. We have since rectified the error by updating the operating parameters of the PLL and repeated measurements for X. Since this excess decay has a resonant feature at Rabi frequencies outside of the regime we operate in X. 
\subsubsection{Survival probability \& SPAM errors}
\label{app:spam}
Finite vacuum limited lifetime of trapped atoms, together with heating due to scattering, can lead to atom losses. These survival probabilities have to be taken into account, irrespective of added noise.  

After accounting for the added noise, survival and Rabi inhomogeneity, SPAM errors (including survival probabilities) are the most important incongruity left between the theoretical and measured spectra. We model these as two separate parameters $p_{0\rightarrow 1}$ and $p_{1\rightarrow 0}$ which respectively encode for the error probability of measuring $|1\rangle$ when the atom was in state $|0\rangle$ and vice versa. The noise-fidelity relations offer a new approach to characterizing SPAM errors by analyzing the discrepancies between the theoretical and measured fidelity spectra. By minimizing the Kullback-Leibler divergence \cite{kullback} between the experimental and simulated fidelity distributions $F_{\text{exp.}}$ and $F_{\text{sim.}}$ at zero noise $D(F_{\text{exp.}}||F_{\text{sim.}})$, one can effectively fit the values of $p_{0\rightarrow 1}$ and $p_{1\rightarrow 0}$ as
\begin{equation}
    (p_{0\rightarrow 1},p_{1\rightarrow 0})=\arg\min_{x\in[0,1]^2} D(F_{\text{exp.}}||F_{x,\text{sim.}})
\end{equation}

These parameters are considered to be array-site independent and are used in the simulated measurement model which we now describe. In total, there are $N_m\cdot N_a$ measurements, with $N_m$ the repetitions of measurements and $N_a$ the array sites, which we index as $l=(j,k)$. Each of the these measurements  is distributed as $M_{i,l}\sim\left(\mathbb{B}_l(p_c),\mathbb{B}_l(F_{ij}),\mathbb{B}_l(p_{0\rightarrow 1}),\mathbb{B}_l(p_{1\rightarrow 0})\right)$ for noise realization $i$ is modeled as the combination of four Bernoulli random variables with probabilities $p_c$, $F_{ij}$, $p_{0\rightarrow 1}$ and $p_{1\rightarrow 0}$ for respectively giving a result, whether that result is a 0 or 1 and (the last two) whether a SPAM error has occurred. If $\mathbb{B}_l(p_c)=0$, the measurement is discarded. If $\mathbb{B}_l(p_c)=1$, the measurement is valid and gives the result $\mathbb{B}_l(F_{ij})$ unless a SPAM error has occurred, which is indicated by $\mathbb{B}_l(p_{0\rightarrow 1}),\mathbb{B}_l(p_{1\rightarrow 0})$, in which case the result is flipped. The simulated measured fidelity $F_{m,i}$ is the average of the measurements that yielded a result,
\begin{equation}
\label{eq:simulatedmeasure}
\begin{aligned}
    &F_{m,i}=\frac{1}{\sum_{l=(j,k)}^{N_mN_A} \mathbb{B}_l(p_c)}\sum_{l=(j,k)}^{N_m\cdot N_a} \mathbb{B}_l(p_c)\\
    \times&\bigg(\mathbb{B}_l(F_{ij})(1-\mathbb{B}_l(p_{1\rightarrow 0})+(1-\mathbb{B}_l(F_{ij}))\mathbb{B}_l(p_{0\rightarrow 1})\bigg)
\end{aligned}
\end{equation}

Note that for large enough $N_m$, simulating for the decisiveness of the experiment via $\mathbb{B}(p_c)$ will not be of great influence on $\mathbb{E}[F_{m,i}]$, but will be important for the entire distribution $F_{m,i}$, producing a smoother and less discrete distribution. For visualization purposes, we will represent this distribution by a kernel density estimation, which takes away the discrete character of the sampled simulation distribution by convolution with a Gaussian kernel. This approximates the true probability density estimate of the fidelities. From this analysis we estimate $p_{0\rightarrow1}\approx 0.04$ and $p_{1\rightarrow0} \approx 0.04$.

%For each experiment, we then match a T-matrix SPAM error to the experimental distribution by minimizing the Kullback-Leibler divergence. This process is described in App.~\ref{app:kldiv}. Including these SPAM errors, measurement on the generated states is simulated 300 times per state (as in the experiment) to result in a simulated fidelity distribution to which the experimental results can be compared fairly. 

\subsubsection{Amplitude Response}
\label{app:ampresp}
\begin{figure}
    \centering
    \includegraphics[width=1.0\linewidth]{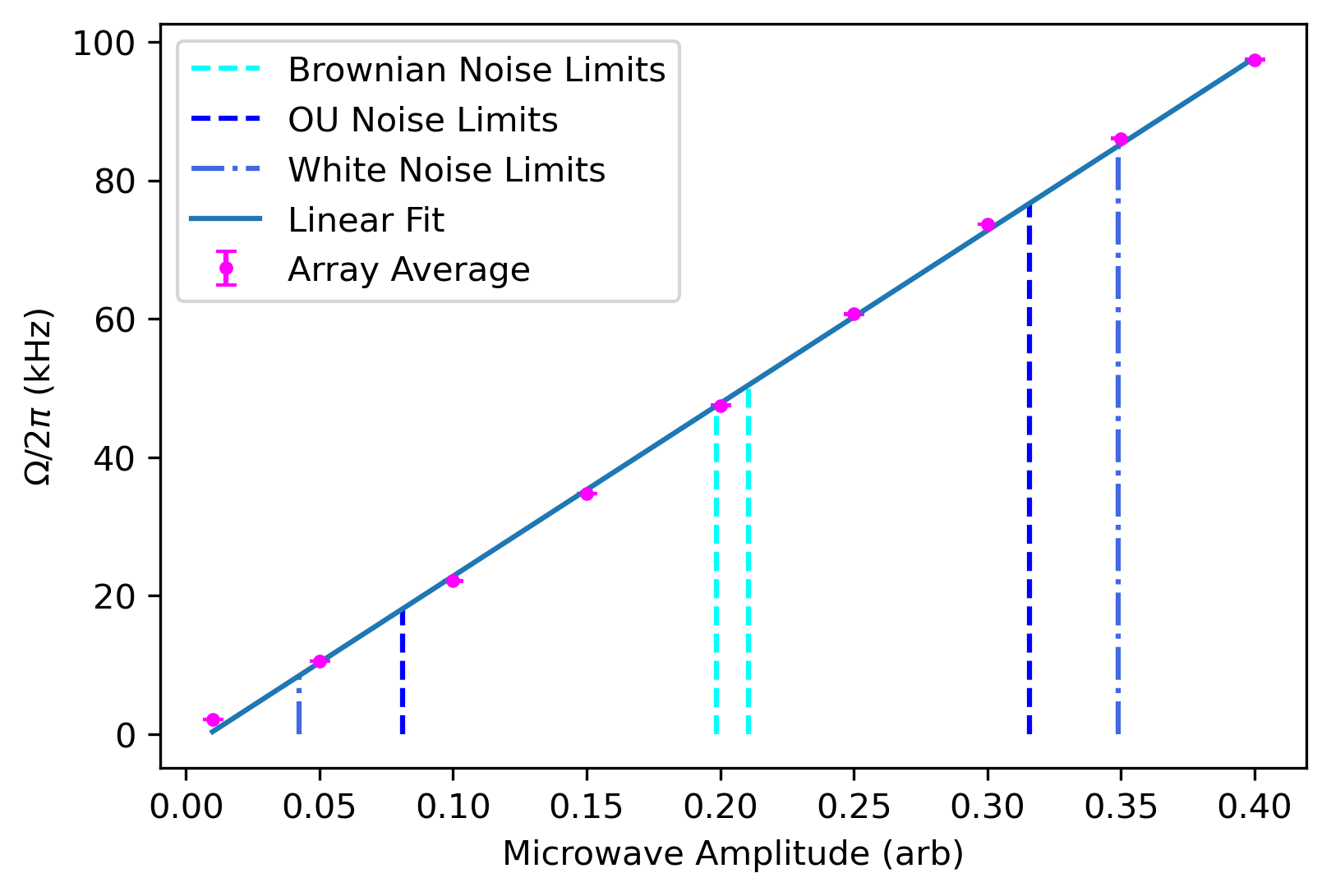}
    \caption{Measured Rabi frequencies for various pulse amplitudes. We indicate the maximum and minimum amplitudes used for the different noise types by the dashed lines. Error bars indicate the uncertainty in the fitted oscillation frequency.}
    \label{fig:RabiLinear}
\end{figure}

To characterize the linearity of the microwave amplitude setpoint and the Rabi frequency experienced by the atoms, we apply a microwave pulse of variable duration and fit a sinusoid to the survival probabilities after the state projection process. We repeat this for various pulse amplitudes. The Rabi frequency has a non-linear response at small amplitudes and linear response in the regime used for artificial noise used throughout the experiment, as depicted in Fig.~\ref{fig:RabiLinear}.
To ensure that the intended pulse is experienced by the atoms, we firstly compare the time-dependent amplitude of microwave signal as measured by the power detector before the antenna with that of the calculated pulse. To further ensure that the intended pulse is applied to the atoms we perform a measurement to determine the maximum rate at which the Rabi frequency can be switched between 0 and $2\pi \times $\SI{50}{\kilo\hertz} without affecting the pulse area experienced by the atoms. We generate a set of pulses, all with duration \SI{200}{\micro\second} of which the Rabi frequency is alternated between $2\pi\times $\SI{50}{\kilo\hertz} and \SI{0}{\kilo\hertz} in segments of duration $\tau$. The effective pulse area is therefore $10\pi$. The number of times the pulse alternates between high and low is chosen as an integer divisor of $50000/2$. We calculate the pulse for these $24$ divisors, which spans the range of $\tau$ between \SI{4}{\nano\second} and \SI{100}{\micro\second}. We measure $P_{\ket{0}}$ for the various pulses and find that for $\tau \geq \SI{400}{\nano\second}$ we measure $<0.006$ deviations in $P_{\ket{0}}$ compared to the $\tau = \SI{100}{\micro\second}$ pulse. Due to the finite slew rate of the microwave amplitude, large deviations (above 0.5) appear at the shortest time interval of \SI{4}{\nano\second}. We use this method to indicate that the minimum duration pulse segment that can be reproduced with confidence is less than the \SI{1}{\micro\second} timestep used throughout the experiment.       

\subsubsection{Additional Error Sources}
\label{app:AddSources} 
Adjusting for spatial Rabi frequency inhomogeneity and SPAM errors greatly improves simulation and experimental overlap, but does not fully remove remaining discrepancies. To mitigate microwave amplitude instabilities, we actively calibrate the microwave amplitude on the atoms approximately every 4 hours, achieving microwave amplitude stability better than $0.2$\% over this timescale. On shorter pulse-length timescales, no significant shot-to-shot fluctuations are observed within the accuracy of the microwave power meter that we use for monitoring. We characterize microwave phase noise with a spectrum analyzer and measure sub-\SI{20}{\hertz} linewidth, indicating minimal error contributions from this source. Tweezer intensity noise and effects of the SLM refresh rate are reduced by lowering the trap depths during all experiments. The trap beam intensity is stabilized before the SLM, and we do observe a small spectral peak at \SI{1.4}{\kilo\hertz} after the SLM, which we attribute to its refresh rate. To suppress magnetic field noise, we trigger the control sequence on the AC line for all our experiments to reduce any \SI{50}{\hertz} noise that may be present. Furthermore, we estimate the magnetic field noise from the T$^*_2$ coherence time of the $5S_{1/2}~\ket{F=2,m_F=1}\leftrightarrow$~$5S_{1/2}~ \ket{F=3,m_F=1}$ transition to $\sigma_B <$ \SI{1.8}{\milli G}.  
\subsection{Noise generation}
\label{app:stepsize}

\begin{figure}
    \centering
    \includegraphics[width=0.99\linewidth]{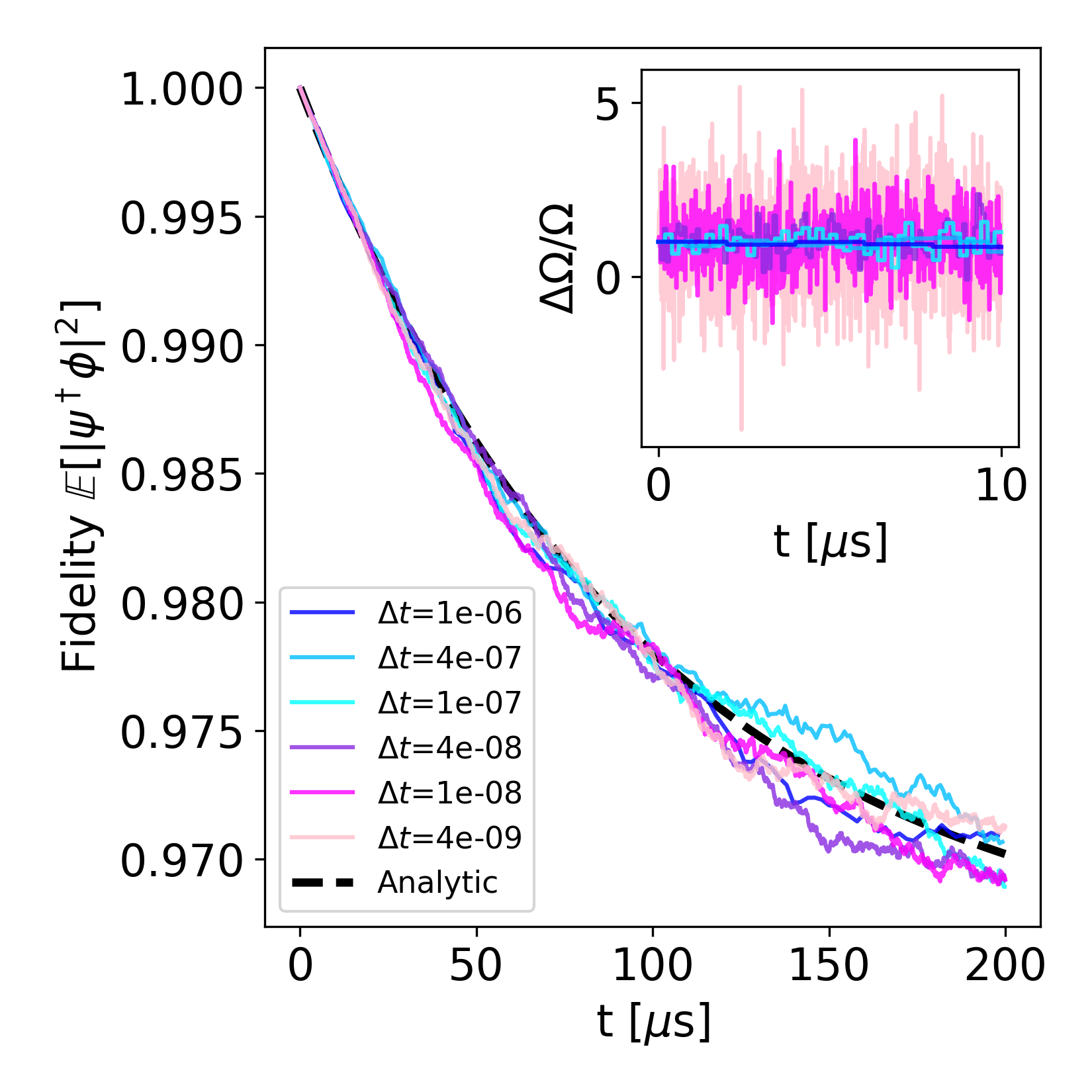}
    \caption{Simulated average fidelity vs. time plot for different timescales of the noise $\Delta t$, together with the analytic result over 200 realizations. Inset: example of noise realization over Rabi frequency.}
    \label{fig:samples}
\end{figure}

\begin{figure}
    \centering
    \includegraphics[width=0.9\linewidth]{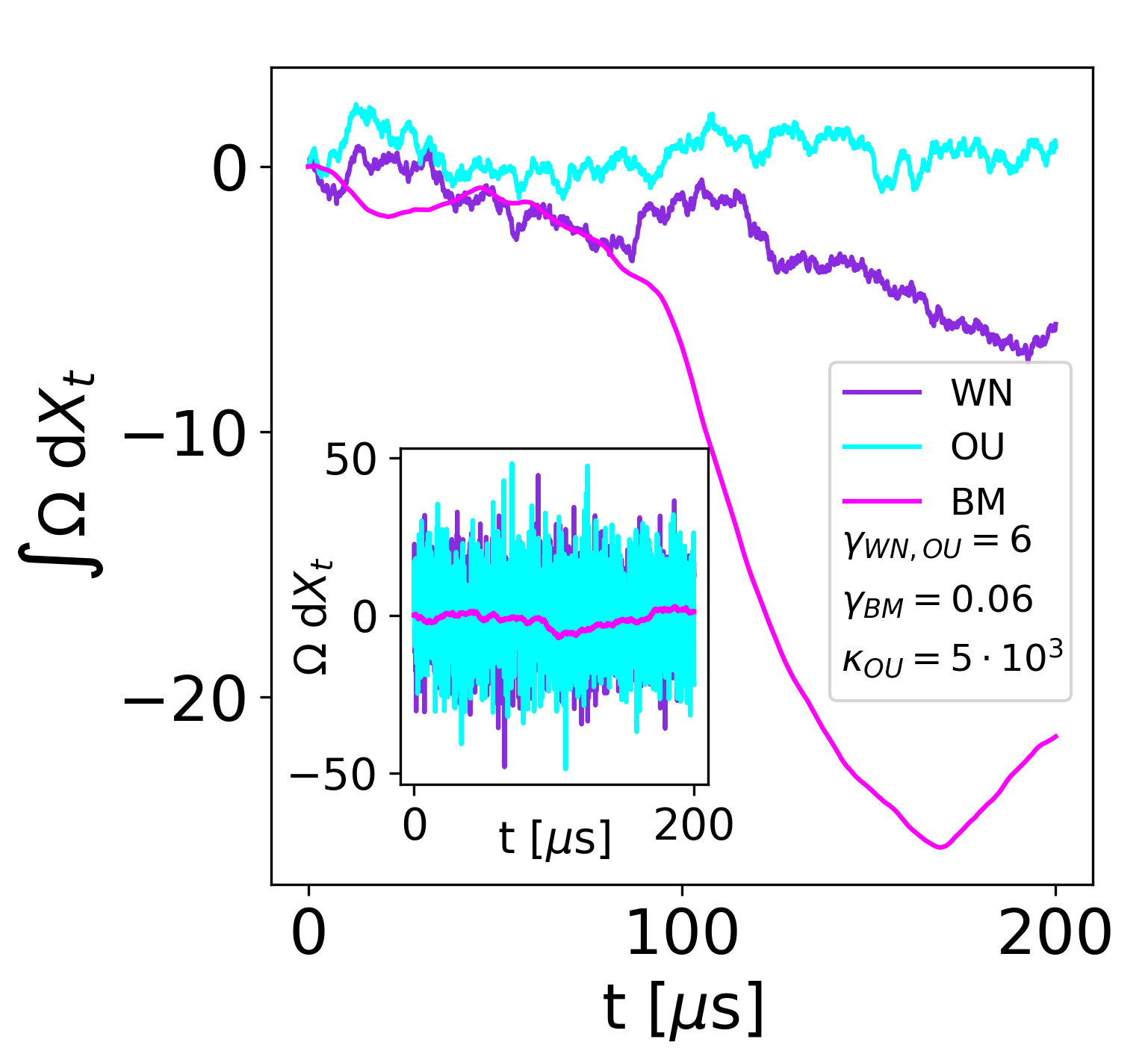}
    \caption{Single noise realizations as a function of pulse duration of white noise (WN), Ornstein-Uhlenbeck (OU) and Brownian motion (BM) with stepsize $\Delta t=\SI{1}{\micro\second}$. Inset: noise increments belonging to main plot as required in Eq.~\eqref{eq:sse}. }
    \label{fig:pulse-examples}
\end{figure}

\begin{figure}
    \centering
    \includegraphics[width=0.9\linewidth]{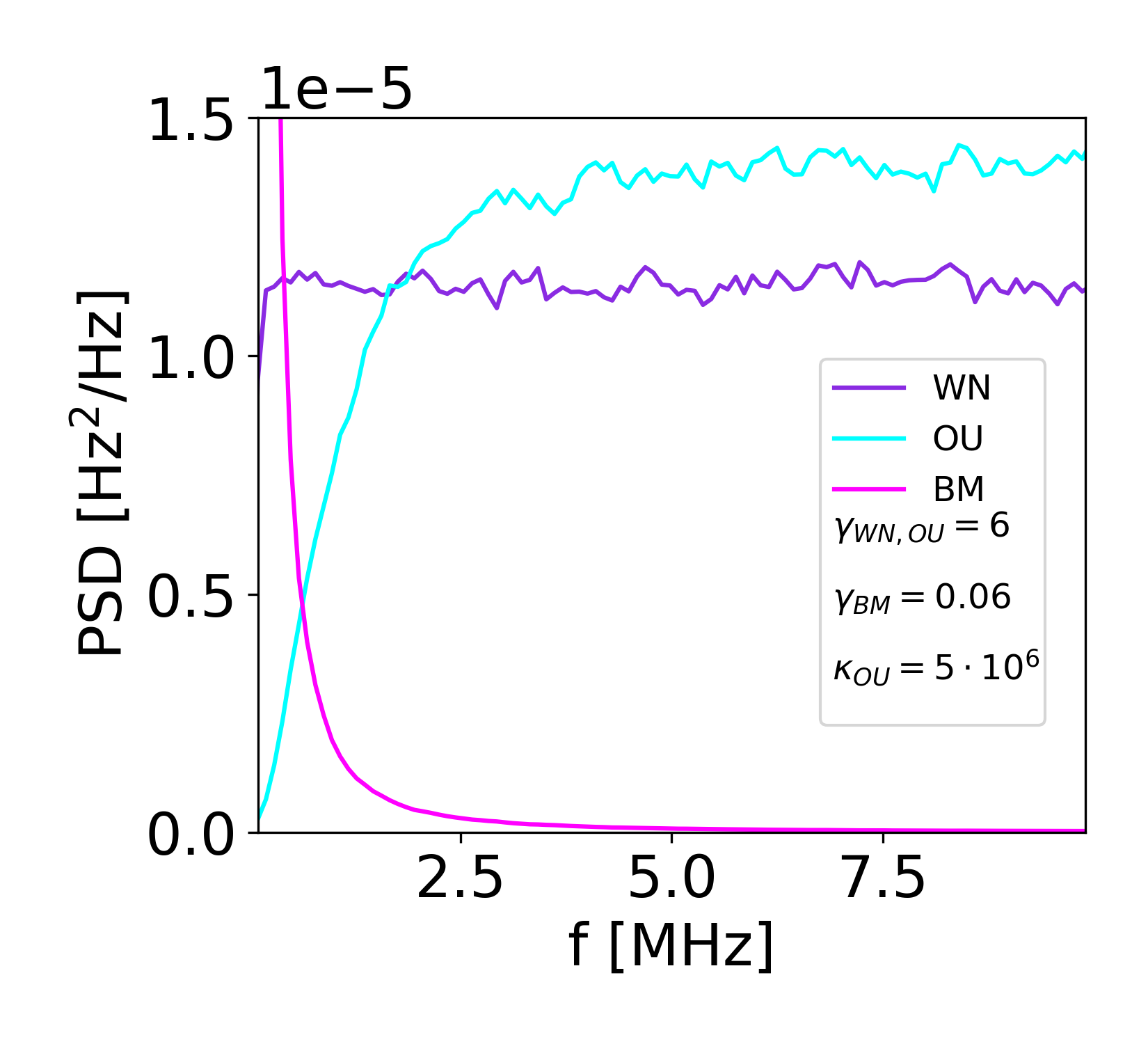}
    \caption{Power spectral densities of white noise (WN), Ornstein-Uhlenbeck (OU) and Brownian motion (BM) with stepsize $\Delta t=\SI{4}{\nano\second}$ and $T=\SI{0.02}{\second}$. $\kappa$ increased to $5\cdot10^6$  s$^{-1}$ for OU to better visualize characteristic PSD. Clear constant PSD for WN, low frequency filter for OU, and $1/f^2$ for BM.}
    \label{fig:psd-examples}
\end{figure}

In the system, we inject three different noise profiles: white noise (WN), Ornstein-Uhlenbeck (OU), and Brownian motion (BM), respectively defined as

\begin{equation}
\label{eq:noises}
\begin{aligned}
    \text{(WN):}\quad & \di X_t= \gamma\,\di W_t\\
	\text{(OU):}\quad &\di X_t=-\kappa X_t\,\di t+\gamma\,\di W_t,\\
    \text{(BM):}\quad & \di X_t = \gamma\gamma_0 \int_{0}^t  \di W_s \di s.
\end{aligned}
\end{equation}
Here $W_t$ is a realization of the Wiener process, which integrated is a Brownian motion \cite{oksendal}. The OU process represents a damped Wiener process with noise strength $\gamma$ and damping rate $\kappa$. These processes model diffusion-relaxation processes, where $\gamma$ is the rate of diffusion and $\kappa$ the rate of relaxation \cite{OUnoise}. $\gamma_0=\SI{1}{\per\second}$ is a unit constant for dimensional consistency. When numerically sampling realizations of this noise as step functions for timesteps of \SI{4}{\nano\second}, excessively large noise values appear due to the incongruence between the noise timescale $\sqrt{\Delta t}$ stemming from the white noise \cite{stochasticintegration1} and the pulse timescale $\Delta t$ (see inset Fig.~\ref{fig:samples}). This would be resolved by refining to a smaller timestep. Instead, we average the noise over several time steps to even out the values, and numerically verify that in simulation these smoothened out noise realizations result in the same fidelity distributions as the \SI{4}{\nano\second} samples, see Fig.~\ref{fig:samples}. In all experimental and simulation results timesteps of $\Delta t=$\SI{1}{\micro\second} are used. Representative noise realizations and power spectral densities (PSD) for each noise profile are respectively shown in Fig.~\ref{fig:pulse-examples} and Fig.~\ref{fig:psd-examples}.

\subsection{Analytic fidelity expressions}
\label{app:analexpres}
For intensity noise, the noise fidelity distribution is given by Eq.~\eqref{eq:fidelity} \cite{ssetheory} 

\begin{equation}
\label{eq:paulidist}
\begin{aligned}
    \F &=\frac{1}{2}(1+\S_0^2)+\frac{1}{2}(1-\S_0^2)\cos(2(X_t-X_0)) \\
    &=\cos^2(X_t-X_0) + \S_0^2\sin^2(X_t-X_0),
\end{aligned}
\end{equation}
where $\S_0=\phi_0^\dagger S \phi_0$. For the (WN), (OU) and (BM) noise processes, explicit formulas for the expectation and variance of the fidelity can be deduced by considering that

\begin{equation}
\label{eq:expcos}
    \mathbb{E}[\cos(\alpha(X_t-X_0))]=\sum_{n=0}^\infty \frac{(-1)^n}{2n!}\alpha^{2n}\mathbb{E}\big[\mathbb{E}[(X_t-X_0)^{2n}|X_0]\big].
\end{equation}

In Ref.~\cite{ssetheory}, recursive differential equations for $(X_t-X_0)^m$ are solved to give closed expressions for the expectations. Resulting in analytic expressions for the mean and variance of the form

\begin{equation}
\label{eq:expectations}
    \mathbb{E}[\F_t]=\begin{cases} \frac{1}{2} \left(1+\S_0^2\right)+\frac{1}{2} \left(1-\S_0^2\right) e^{-2\gamma^2 t},\quad\quad \text {(WN)}\\
    \frac{1}{2} \left(1+\S_0^2\right)+\frac{1}{2} \left(1-\S_0^2\right) e^{-2\gamma^2\tau_\kappa(t)}, \quad \text{(OU)}\\ 
    \frac{1}{2} \left(1+\S_0^2\right)+\frac{1}{2} \left(1-\S_0^2\right) e^{-\frac{2}{3}\gamma_0^2\gamma^2t^3},\quad\quad \text{(BM)}\\
\end{cases}
\end{equation}

\begin{equation}
\label{eq:variances}
     \text{Var}(\F_t)=\begin{cases} \frac{1}{8} \left(1-\S_0^2\right)^2 \left[1-e^{-4\gamma^2t}\right]^2,\quad\quad \text {(WN)}\\
    \frac{1}{8} \left(1-\S_0^2\right)^2 \left[1-e^{-4\gamma^2\tau_\kappa(t)}\right]^2, \quad \text{(OU)}\\ 
    \frac{1}{8} \left(1-\S_0^2\right)^2 \left[1-e^{-\frac{2}{3}\gamma_0^2\gamma^2t^3}\right]^2,\quad\quad \text{(BM)}\\
\end{cases}
\end{equation}

with $\tau_\kappa(t) := e^{-\kappa t} \sinh (\kappa t)/{\kappa}$.

\subsection{Stochastic Integration}
\label{app:stochint}
Numerical stochastic integration is performed by solving for the noise and state simultaneously, we define $\Y:=(\psi,X)$. For the (OU) process, this gives the differential equation
\begin{equation}
\begin{aligned}
    &\di \Y=a(\Y)\,\di t+b(\Y)\,\di W_t,\\
    a(\Y)=&
\begin{pmatrix}
-iH+ikXS-\frac{\gamma^2}{2}S^\dagger S
  & \rvline & \mathbf{0} \\
\hline
  \mathbf{0}^\text{T} & \rvline &
-k
\end{pmatrix}\Y,\\
&b(\Y)=\begin{pmatrix}
-i\gamma S
  & \rvline & \mathbf{0} \\
\hline
  \mathbf{0}^\text{T} & \rvline &
\frac{1}{X}
\end{pmatrix}\Y.
\end{aligned}
\end{equation}
These equations can be solved discretely over time steps $\Delta t$ using a numerical integration scheme. The non-Lipschitz $1/X$ dependence \cite{platen} and possibly the non-Euclidean space in which the states live can cause convergence issues \cite{ssetheory}. These convergence issues are absent for white noise and always occur below fidelities of $F=0.95$, which is not a relevant regime for pragmatic quantum computing. These problems are mitigated (but not resolved) by using a higher-order scheme such as the explicit (weak) second-order Platen scheme \cite{stochasticintegration2,platen}, which is used for simulation in this work. The scheme is
$$
\begin{aligned}
\Y_{n+1}=&\Y_n+\frac{1}{2}\big(a(\bar{\Upsilon})+a(\Y_n)\big) \Delta t\\
&+\frac{1}{4}\big(b\left(\bar{\Upsilon}^{+}\right)+b\left(\bar{\Upsilon}^{-}\right)+2 b(\Y_n)\big) \mathcal{N}\sqrt{\Delta t}\\
&+\frac{1}{4}\big(b(\tilde{\Upsilon}^{+})-b(\widetilde{\Upsilon}^{-})\big)\left(\mathcal{N}^2-1\right) \sqrt{\Delta t}, 
\end{aligned}
$$
with supporting values $\bar{\Upsilon}=\Y_n+a(\Y_n) \Delta t+b(\Y_n) \mathcal{N}\sqrt{\Delta t}$ and $\bar{\Upsilon}^{ \pm}=\Y_n+a(\Y_n) \Delta t \pm b(\Y_n) \sqrt{\Delta t}$.

\appendix

\end{document}